\documentstyle[12pt]{article}
\def\beq{\begin{equation}}
\def\eeq{\end{equation}}
\def\bea{\begin{eqnarray}}
\def\eea{\end{eqnarray}}
\def\bq{\begin{quote}}
\def\eq{\end{quote}}

\def\beqa{\begin{eqnarray}} 
\def\eeqa{\end{eqnarray}} 
\def\be{\begin{equation}} 
\def\ee{\end{equation}} 
\def\beq{\begin{equation}}   
\def\eeq{\end{equation}}

\def\pa{\partial} 
\def\kaps{{\kappa}^{2}} 
\def\Melev{ M^{11}}

\def\bi{\begin{itemize}} 
\def\ei{\end{itemize}}

\def\gappeq{\mathrel{\rlap
{\raise.5ex\hbox{$>$}}
{\lower.5ex\hbox{$\sim$}}}}

\def\lappeq{\mathrel{\rlap{\raise.5ex\hbox{$<$}}
{\lower.5ex\hbox{$\sim$}}}}

\begin{document}
\pagestyle{empty}
\begin{flushright}
{
CERN-TH/98-230\\
hep-ph/9807503}
\end{flushright}
\vspace*{5mm}
\begin{center}
{\bf 
Beyond the Standard Embedding in 
$M$-Theory on $S^{1}/Z_{2}$ 
} \\
\vspace*{1cm} 
Zygmunt Lalak$^{a,b)}$, Stefan Pokorski$^{a,b)}$, Steven Thomas $^{c)}$ 
{ }\\
\vspace{0.3cm}
\vspace*{2cm}  
{\bf ABSTRACT} \\
\end{center}
\vspace*{5mm}
\noindent
In this paper we discuss  compactifications  of  M-theory  
to four dimensions on $X \times S^{1}/Z_{2} $, 
in which  nonstandard embeddings  in the
$E_8 \times E_8 $ vacuum  gauge bundle are considered. At the level of 
the effective field theory description of Horava and Witten, this 
provides  a natural extension of  well known results at weak coupling, to strongly coupled  $E_8 \times E_8 $ heterotic strings. As an application of 
our results, we discuss  models  which exhibit  an anomalous $U(1)_A$ symmetry
in four dimensions, and show how this emerges from the reduction of the
 $d = 11 $ toplogical term $C \wedge G \wedge G $, and how it is consistent
 with $d = 4 $  anomaly cancellation in M-theory.  As a further application of 
nonstandard embeddings, we show how it is possible  to obtain 
an inverse hierarchy of gauge couplings, where the observable 
sector is more strongly coupled than the hidden one. 
The basic construction and phenomenological viability of these scenarios 
is demonstrated. 
 
\vspace*{1cm} 
\noindent

\rule[.1in]{16.5cm}{.002in}

\noindent
$^{a)}$ Theory Division, CERN, Geneva, Switzerland.\\
$^{b)}$ Institute of Theoretical Physics, Warsaw University.\\
$^{c)}$ Department of Physics, Queen Mary and Westfield College, London. \\
\vspace*{0.5cm}
\begin{flushleft} 
CERN-TH/98-230\\
July 1998
\end{flushleft}
\vfill\eject

\setcounter{page}{1}
\pagestyle{plain}

\section{Introduction}

The quest for the theory unifying all the known interactions
has entered a qualitatively new and exciting phase 
with the realization of the existence of the web of dualities connecting 
apparently different string models and allowing for the exploration of 
nonperturbative phenomena in string theory. 
The underlying eleven dimensional quantum theory whose ten dimensional 
emanations are known superstring theories has been termed $M$-theory.
One simple-minded way of viewing the relation of M-theory to strings is  
to say that in this framework the string coupling
becomes a dynamical field, which may be interpreted as an
extra spatial dimension. Many field theoretical limits corresponding to 
compactifications of M-theory on various manifolds and orbifolds are known.
Among them special role is played by the chiral $N=1$ supersymmetric model 
formulated by Horava and Witten, which is the low-energy limit of the 
$M$-theory compactified on the line segment $S^{1}/Z_2$ and further 
on a six dimensional Calabi-Yau manifold \cite{wh1,strw,wh}. 
The field theoretical 
model is constructed 
as a consistent compactification of the eleven dimensional $N=1$ 
supergravity theory on $S^{1}/Z_2$. It has been 
demonstrated that the supergravity model obtained this way, living on the 
eleven dimensional manifold with two parts 
of its ten dimensional 
boundary inhabited by the ten dimensional $E_8$ gauge supermultiplets,  
forms the low-energy limit of the strongly coupled
$E_8 \, \times \, E_8$ heterotic superstring \cite{wh1}.  
One of the important results of this construction is that, in order to
preserve
 supersymmetry and to assure the absence of anomalies, the usual 
stringy Bianchi identity has to be modified \cite{strw}. 
This results in a nonvanishing 
antisymmetric tensor field background. As a consequence, corrections to the 
Calabi-Yau metric are induced. Witten has shown that the volume of the C-Y space 
changes linearly with $x^{11}$ along the eleventh dimension, with 
a coefficient which depends on the gauge and gravitational 
vacuum configurations. 
The important consequence for physics in four dimensions is that the 
gauge couplings in the hidden and visible sectors are split with the sign of 
$\delta (\frac{1}{\alpha})\, = \, \alpha_h^{-1} - \alpha_v^{-1} $
 and its magnitude depending on the particular 
embedding of the gauge vacuum bundle in the $E_8 \times E_8$ bundle. 
For the standard embedding, which consists in identifying the 
$SU(3)$ holonomy gauge field of the underlying Calabi-Yau manifold with 
some of the gauge fields 
from one of the two $E_8$s - the one which is subsequently broken down to
 $E_6$ and  called the visible gauge sector,
 $\alpha_v \, < \, \alpha_h$. Here indices $v,\; \; h$ denote the visible 
and hidden sectors and $\alpha = g^2 /(4 \pi) $. 

Moreover,  additional consistency checks appear,  namely 
 to have $\alpha_h \, < \, \infty$,
the physical radius of the eleventh dimension, $R_{11}$, must be smaller 
than 
certain critical value $R_{crit}$.  On the other hand, to obtain the 
correct value of the Newton constant, $G_N$, we need 
$R_{11} \approx R_{crit}$. Therefore, $\alpha_{h}$ at $M_{GUT}$ is 
relatively large,
and this
 raises the question about the scale of the gaugino 
condensate, and about visible mass hierarchy which should follow (there seems 
to be no room left for running of the hidden gauge coupling below $M_{GUT}$). 

This problem seems to be directly related to the fact that, although 
nonstandard embeddings were already discussed in \cite{min}, 
the examples of the specific $M$-theoretical model considered so far in the 
literature have been obtained precisely in the context of the standard 
embedding \cite{msus}. It is an obvious idea that relaxing the standard embedding 
assumption might help to obtain a more realistic scale of the
 gaugino condensation 
in $M$-theory. 
Moreover, as we shall see, the standard embedding is consistent with the 
overall unification of couplings only with $\alpha_v \, < \, 
 4 \pi^{3/2} \frac{1}{\epsilon^{3/4}}
 (M_{GUT} /M_{Pl})^{3/2} $ where $\epsilon$ is a parameter which cannot be 
too small (see Section 5 )\footnote{Models with $\epsilon = 0$ exist, but are phenomenologically uninteresting.}.  Thus, typically, there is
no room left for the possibility of unification at `strong' coupling, 
which is characteristic for models with additional matter 
at intermediate scales 
\cite{models}. 
Again, relaxing the standard embedding assumption opens up this 
avenue (see also \cite{benakli}). One should stress 
that both problems are typical for $M$-theory. 
In the weakly coupled heterotic string, although there exist similar 
correlations between the gauge couplings of the two $E_8$s, the difference 
$\delta (\frac{1}{\alpha })$ is generically small and phenomenologically 
irrelevant. 

The above genuinely $M$-theoretical motivation for going beyond the standard
embedding comes
 in addition to the reasons which are the same as for the weakly coupled string models \cite{gsw,dist,kach} {}\footnote{A possibility of extending 
nonstandard stringy models to $M$-theory was considered in \cite{far}.}. 
As in the weakly coupled case, these constructions 
give a chance of constructing directly models similar to the MSSM, 
with non-semisimple and lower rank gauge group at the unification scale. 
Also,
this seems to open up the possibility of  gauge mediated 
supersymmetry transmission 
models \cite{gmod} in the context of the Horava-Witten model. 
Another very important fact is that, as in the weakly coupled case, 
the anomalous $U(1)_A$
group can be present only in models which go beyond the standard 
 embedding. 
The physics of models with the anomalous  $U(1)_A$ has been studied 
extensively over the last years and it has been shown 
that it can be 
relevant for the fermion mass generation \cite{ross}, 
and 
for the supersymmetry breaking issue \cite{b7,b12,lalak}. Hence, it is of 
considerable 
importance to set up the framework for the discussion of 
the anomalous  $U(1)_A$ symmetry in the context of the strongly 
coupled heterotic string models. 

The purpose of this paper is to give a setup for the Horava-Witten 
model with non-standard embeddings and to discuss some of the 
 phenomenological questions raised above.

Let us summarize 
here some basic facts 
worked out in the literature so far. 
First, we note  that the calculation of Witten, which gives the picture 
of unification in $M$-theory, is valid for general embeddings. 
However, it gives only  $\delta (\frac{1}{\alpha})$, not the 
normalization of the
individual couplings separately, and it has been performed in the eleven 
dimensional setup, in particular using eleven dimensional metric and eleven 
dimensional equations of motion. 
Ultimately, we are interested in the four dimensional effective theory
and would like, for instance, to express the gauge couplings in each 
gauge sector in terms of the four dimensional moduli chiral superfields
$S$ and $T$. In particular, supersymmetry breaking may be governed by the 
four dimensional physics. Moreover, the comparison to the weakly coupled heterotic string predictions is transparent in four dimensions. 

The correct procedure for obtaining the effective four dimensional theory 
consists in integrating out physics in six dimensions compactified on the 
Calabi-Yau manifold and in the eleventh dimension compactified 
on the $S^1 /Z_2$ \cite{hpn1,lt,pesk,low5,elpp}.
This should be performed in this well defined order as, for
 phenomenological reasons, the eleventh dimension has to be about 
an order of magnitude larger than the Calabi-Yau radius. 
Integrating out the six C-Y dimensions introduces the Calabi-Yau modes 
into the (nonlinear sigma model) five dimensional supergravity Lagrangian,
which in a sense is the fundamental Lagrangian of the $M$-theory. 
It is in five dimensions where we should solve the equations of motion 
with the sources defined with the $S^1 / Z_2$ compactification, and afterwards 
integrate out the fifth dimension \cite{low5,elpp}. This approach is 
considerably 
complex and has not yet been fully completed. A simplification which has been 
taken in several papers (including the original paper \cite{strw}) 
and which we assume here is to neglect the effects of the Calabi-Yau physics 
on the dynamics in the fifth dimension. This way the equations of motion in the fifth (eleventh) dimension can be solved directly along the
 eleventh dimension. After dimensional reduction and truncation of the six C-Y
dimensions, and Kaluza-Klein reduction from five to four dimensions, the effective 
four dimensional theory is obtained. 

Returning now to the existing literature,
it  was the work of 
Banks and Dine \cite{bd} which has shown that with the standard embedding, 
Witten's result  for $\delta (\frac{1}{\alpha})$ follows from the standard
universal threshold corrections $\frac{1}{g^{2}_{(1),(2)}} = S_r 
\pm \epsilon
 T_r $ derived 
in the weakly coupled heterotic string  
from the Green-Schwarz terms in ten dimensions (for earlier work in this direction see Ibanez and Nilles \cite{iba}, axionic couplings were also considered
in \cite{chk}). The analysis of the relation between 
weakly coupled string one-loop threshold corrections and strongly coupled 
string threshold corrections has been performed also in \cite{sstieb},
and recently threshold corrections for nonstandard embeddings 
have been computed in \cite{step}. 
Then Lukas {\em et al.} \cite{low10}
have shown that the Green-Schwarz terms used in \cite{bd} can be derived from 
the $11d\, \to \, 10d$ Kaluza-Klein reduction,
 in the spirit of the simplified approach discussed above. 
Hence, it has been shown that 
the overall normalization of the Witten's result follows from the 
axionic couplings, and actually for the standard embedding the 
effective four dimensional form $1/ g^{2}_{v,h} = S_r \pm \epsilon T_r$ 
is derived from the eleven dimensional theory, and it agrees with the 
explicit weak coupling calculation. 
This is very interesting since, in the context of the five dimensional physics,
the origin of axionic corrections and scalar corrections 
is totally different (the former come through the topological interaction term 
in eleven or five dimensions, and the latter through the corrections to the 
C-Y metric). 
It confirms the fact that, at least for the gauge fields and charged matter 
sector,
 there should exist a reduced effective theory which has the form of the four dimensional supersymmetric gauge chiral model. 

In this paper we generalize the standard embedding result and 
 confirm that the general result
 $ 1/ g^{2}_{(1),(2)} = Re(S + (\pm \epsilon_i) T_i )$ 
(where $i=1,\ldots,h{(1,1)}$) 
is correct for a general embedding if one considers only zero-modes both on 
C-Y space and on the circle. 
In this derivation we follow the simplified 
approach outlined above to the construction of the effective four dimensional theory 
with, however, proper care taken of non-zero modes both on the Calabi-Yau
 space and on the circle in the eleventh dimension. The overall
 normalization
 of gauge couplings is fixed by the normalization of axionic corrections for general 
embeddings. 
Then, we consider the existence of an  
anomalous $U(1)_A$ in the non-standard embedding Horava-Witten model,
Finally, we discuss the possibilities for 
changing the relative sign of threshold  corrections between the different
gauge group factors, a question which is crucial in understanding  gaugino 
condensation hierarchy and 'strong coupling' unification 
in the context of the most general 
embeddings.

\section{Effective Lagrangian of the M-Theory compactified on $S^1 /Z_2$}

To start with let us recall the form of the M-theory Lagrangian 
constructed by Horava and Witten \cite{strw,wh}, which is 
given by $L_S + L_B
$ where 
\beqa\label{e:1} 
 {L_S} & = & \frac{1}{\kappa^2}\int_{\Melev} d^{11}x \, \sqrt{g} \, \{ 
 -\frac{1}{2} R - \frac{1}{2} {\bar{\Psi}}_I \Gamma^{IJK} D_J  {\Psi}_K 
 - \frac{1}{48} G_{IJKL} G^{IJKL} \cr 
&&\cr 
& - & \frac{\sqrt{2}}{384} ( 
 {\bar{\Psi}}_I \Gamma^{IJKLMN} \Psi_N + 12 {\bar{\Psi}}^J \Gamma^{KL} \Psi^M ) 
\,( G_{JKLM} + {\hat{G}}_{JKLM} ) \cr  
&&\cr 
 &-& \frac{\sqrt{2}}{3456} \epsilon^{I_1...I_{11}} C_{{I_1}{I_2}{I_3}} 
  G_{I_4 ...I_7} G_{I_8 ...I _{11}} \}
\eeqa 
 
\beqa\label{e:2} 
 {L_B} & = &  \frac{1}{2 \pi {(4 \pi \kaps )}^{2/3} }   \sum_{m=1}^2 
 \int_{M^{10}_m} d^{10} x  \, \sqrt{g} \, \{ -\frac{1}{4} {\rm Tr}\, 
F^{m}_{AB} 
F^{mAB} - \frac{1}{2} {\rm Tr}\, \bar{\chi}^{m} \Gamma^A D_A ({\hat{\Omega}
} )\, 
\chi^{m} \cr 
&&\cr 
 &- & \frac{1}{8} {\rm Tr}\, {\bar{\Psi}}_A \Gamma^{BC} \Gamma^A 
 (F^{m}_{BC} + {\hat{F}}^{m}_{BC} )\, \chi^{m} + 
 \frac{\sqrt{2}}{48}{\rm Tr}\, {\bar{\chi}}^{m} \Gamma^{ABC} \chi^{m} 
 \, \hat{G}_{ABC11} \} \cr 
&&\cr 
 &+&  O(\kappa^{4/3}) \; {\rm relative}\; { \rm to}\; L_S   
\eeqa 
where in eqs(\ref{e:1},\ref{e:2}),  
$I = 1, ..11 $ label coordinates on $ M_{11} $; 
 $ A= 1..10 $ those on $M_4 \times X $, ( $a, \bar{b} = 1 ..3$ 
will denote the  
holomorphic and antiholomorphic coordinates on $X$.); 
$\kappa = m_{11}^{-9/2} $, with
$m_{11} $ the 11 dimensional Planck mass.  
  The field strength $G_{IJKL}=[ 
 \partial_I C_{JKL} \pm 23 ]\;{\rm terms} + O(\kaps) $ satisfy 
 the modified Bianchi identities \cite{strw} 
 \beq\label{e:5} 
 (dG)_{11ABCD} = -\frac{3 \sqrt{2}{\kaps}}{\lambda^2} 
 \sum_{m=1}^2 \delta^{(m)} (x^{11})\, [ {\rm tr}F^{m}_{[AB}\,F^{m}_{CD]} 
 - \frac{1}{2} {\rm tr} R_{[AB} \, R_{CD]} ] 
 \eeq 
where $\lambda^2 = 2 \pi (4 \pi \kaps )^{2/3} $ is 
the $d=10$ gauge coupling constant.
 The delta functions $\delta^{(m)} (x^{11}) $ 
have support on the two fixed point sets in 
$M_4\times X \times {S^1}/ {Z_2} $. 
The presence of these various source terms in  eq(\ref{e:5}) 
are an important difference with the corresponding Bianchi identities 
relevant to compactification of the perturbative $E_8 \times E_8 $ 
heterotic string, where $H_{ABC} $ plays the role of $G_{11ABC} $. 
 
In the bulk $d=11$ supergravity lagrangian $L_S$, $g = {\rm det} (g_{IJ})$ 
involves the $d=11$ bulk metric. In the boundary lagrangian, 
the same quantity is understood as being the determinant of the 
$d=10 $ metric obtained as the restriction of the bulk metric to 
 either of the two boundaries $M^{10}_m, m = 1,2 $. Similarly the 
 two copies of the $E_8 $ super Yang Mills fields defined on the boundaries 
 are denoted by $F^{mAB} , \chi^{m} $ respectively. $\Omega_{ABC} $ are the 
 usual $d=10$ spin connections, with hatted quantities denoting the 
 supercovariant generalizations, explicit definitions of which 
 can be found in \cite{strw,wh}. 
 
\subsection{Lagrangian at Order $O(\kappa^{2/3})$ and Unification}
                                                      
One of the reasons why the supergravity on a manifold with 
boundary \cite{wh} 
which is the field theoretical limit of the strongly coupled $E_8 \times E_8$
heterotic string is so interesting is that it is possible with its three 
independent low energy parameters to fit three fundamental low energy 
observables in an internally consistent way. Actually, fitting the fundamental 
parameters of the weakly coupled heterotic string (also three of them) 
to observed values of the Newton constant, $G_N$, $g^{2}_{GUT}$ and $M_{GUT}$ 
is also possible, but the value of the dimensionless string coupling
which corresponds to that fit is much larger than unity, signaling 
departure from the perturbative regime. When such a matching is performed 
within the scope of the Horava-Witten Lagrangian, the corresponding string coupling turns out to be large too, but here it is consistent with the assumptions under which the effective field theory limit has been obtained. 

Below we shall briefly review the weakly and strongly coupled string unifications, forgetting in this section about the parts of the Horava-Witten Lagrangian 
which of the order $O(\kappa^{4/3})$ relative to the gravitational part. 
One of the reasons is to establish a clear correspondence between the 
weakly coupled string, and strongly coupled string degrees of freedom, 
which is necessary if one wants to use certain results, like specific values 
of threshold corrections, obtained in the weakly coupled region, in the 
Horava-Witten theory. The correspondence is also useful when one wants 
to realize why the matching conditions are different in both cases, and
 what is the physical interpretation of this difference. 

Let us start with the weakly coupled string. The relevant terms are 
\beq
- \frac{1}{2 \kappa_{10}^{2}}  \int \sqrt{g^{(10)}} R^{(10)} 
- \frac{1}{4 \lambda^{(10) \, 2}} \int \sqrt{g^{(10)}} \phi^{-3/4} \, F^2
\eeq
where the ten dimensional gauge coupling $\lambda^{(10) \, 2}$ can be 
made equal to one through the rescaling of the ten dimensional 
dilaton field $\phi$. It is instructive to recall, that the
ten dimensional supergravity Lagrangian can be obtained as a reduction of the 
eleven dimensional supergravity Lagrangian. In terms of the 
underlying 11d metric $g_{IJ}$ the dilaton is just $\phi^2 = g_{11 \; 11}$.
After truncation the canonical normalization of the 10d Einstein-Hilbert 
is achieved through the suitable Weyl rescaling of the 11d metric,
$g_{MN} \to \phi^{-1/4} g_{MN}$. There is one more useful observation. 
When in the 10d Lagrangian one performs the Weyl rescaling $g_{MN} \to
\phi^{-1} g_{MN}$, then the dilaton $\phi$ enters the Lagrangian only as an overall 
factor, multiplying all terms in the Lagrangian, namely as $\phi^{-3}$. 
This agrees with the effective Lagrangian of the heterotic string if 
one identifies the dimensional string coupling as $g_s = \phi^{3/2}$. 
After these clarifications, let us go back to the canonical 10d Lagrangian, 
and let us perform further reduction down to d=4. Let us write the 10d metric in the form:
\beq
g^{(10)}_{\mu \nu} = e^{-3 \sigma} g^{(4)}_{\mu \nu}, \; \;
g^{(10)}_{M N}= e^{\sigma} g^{(0)}_{M N}
\eeq
where the $g^{(0)}$ is the reference metric on the C-Y space given in the 
context of the canonical metric in $d=10$. 
The volume of the Calabi-Yau space is defined as $V_{X \,w} = e^{3 \sigma}
 \int_X \sqrt{det{ g^{(0)}_{M N}}}$ and $w$ stands for weak coupling. 
Now, when we compactify to four dimensions, then it turns out that 
the 4d metric $g^{(4)}_{\mu \nu}$ defined above is canonical,
and one can easily read off the 4d Planck scale, and 4d gauge coupling:
\beq
\frac{M_{Pl}^{(4) \, 2}}{16 \pi} =\frac{M_{Pl}^{(10) \, 8}}{16 \pi} \, V^{(0)}_w,
\; \; \frac{1}{g^{2}_{GUT}} = \frac{V^{(0)}_w}{ \lambda^{(10) \, 2}} e^{3 \sigma} \phi^{-3/4}
\eeq
In the above there are three stringy parameters, $ M_{Pl}^{(10)}, \; \phi, \; 
\sigma$, and two reference constants which we are free to choose in any convenient way. The standard choice is $V^{(0)}_w = M_{Pl}^{(10)\, -6}$ 
which makes $M_{Pl}^{(10)} = M_{Pl}^{(4)}$, and $\frac{V^{(0)}_w}{ \lambda^{(10) \, 2}} =1$. In addition, one has to express the four dimensional 
unification scale $M_{GUT}$ in terms of the independent parameters.
The natural choice is $M_{GUT}^{-6} = V_{X \,w} = e^{3 \sigma} M_{Pl}^{(10)\,
 -6}$. In the context of the weak string Lagrangian in four dimensions
one introduces chiral moduli superfields $S_w , \; T_w$ whose real parts are
$ S_{w \, r} = e^{3 \sigma} \phi^{-3/4}$ and $T_{w \, r} = e^{\sigma}
\phi^{3/4}$. If we take the  values of observables to be 
$g_{GUT} = 0.7, \; M_{Pl} = 1.2 \times 10^{19} \, GeV, \; M_{GUT} =2 \times 10^{16} \,
 GeV$, then we obtain from $S_{wr} = 1/ g^{2}_{GUT}, \; 
 S_{wr} T_{wr} = (M_{Pl}/ M_{GUT})^{8}$ 
the values $S_{wr} = 2.04$ and $T_{wr} = 8 \times 10^{21}$ 
and the corresponding value of the string coupling $g^{-2}_{s} = 
S_{wr}/T_{wr}^{3} =   4 \times 10^{-66}$ which is clearly inconsistent 
with the assumption of the perturbativity of the underlying string theory. 
Such huge values of $T_{wr} $ are just symptomatic of the failure of isotropic
C-Y spaces to 'decouple' the scales  $M_{Pl} $ from $M_{GUT} $.  
As has been pointed out in \cite{strw},  a way of ameliorating this difficulty is to advocate anisotropic C-Y spaces where $V_w \sim {\alpha' }^{d/2} \, {M_{GUT}}^{d-6} $.
In such a scheme it may be feasible to fit correctly the values of $M_{Pl} , M_{GUT} $ and $ g^2_{GUT} $ and obtain the more reasonable values of $T_{wr} $ that have been used in the past literature, when discussing, for example the role of moduli dependent threshold corrections to the gauge couplings.
Such a solution could hardly be described as natural, however.

The above picture of stringy unification in the weakly coupled regime
can be supplemented by the additional requirement of numerical 
unification of the gauge couplings and 
the dimensionless gravitational coupling $g_{grav}^{2}(E)=
16 \pi G_N E^2$ \cite{polch}. The observation which leads us to 
contemplate this possibility 
is that in ten dimensional {\em stringy} frame one has the relation
$g^{2}_{GUT}\, = \, 16 \pi G_N / \alpha' \, = \,16 \pi G_N M^{2}_{string}$. 
A way to state that this universal unification does not work 
is to assume that all forces numerically unify at 
the observed scale of unification of gauge forces with the unified 
coupling at that scale equal to $g_{GUT}$. Then one easily obtains from the 
above relation 
the value of $M_{pl}$: $ M_{pl} = \sqrt{16 \pi } M_{GUT} / g_{GUT} \approx 
2 \times 10^{17} \; GeV$ which is too small, by two orders of magnitude.  
If one goes to the four dimensional canonical frame, the equivalent 
statement is simply that $1/ \alpha' \, = \, M_{Pl}^{2} / (16 \pi ) $
which is different from $M_{GUT}^{2}$.  

Let us move now to the case of the strongly coupled heterotic string. 
In the analysis presented in this section we shall neglect all the terms 
in the Horava-Witten Lagrangian which are beyond the order $O(\kappa^{2/3})$
relative to the gravitational Lagrangian. 

In this case we start with the eleven dimensional Lagrangian, and 
in first step reduce it directly to five dimensions. As argued in 
\cite{low5,elpp} 
it is five dimensions where the low energy Lagrangian of M-theory 
is naturally formulated, and where the presumed unification of gauge 
couplings with gravity might occur.  

The reduction is made through the ansatz 
\beq \label{e:fr}
g^{(11)}_{\alpha \beta} = e^{-2 \beta} g^{(5)}_{\alpha \gamma}, \; \;
g^{(11)}_{M N}= e^{\beta} g^{(0)}_{M N}
\eeq
where $\alpha,\, \gamma$ are five dimensional indices. We note that
the volume of the Calabi-Yau space is defined now as $V_{X \, s}
= e^{3 \beta } V^{(0)}_{s} = e^{3 \beta }  \int_X \sqrt{det{ g^{(0)}_{M N}}}$ 
in terms of the underlying eleven dimensional metric ($s$ stands for strong 
coupling). In five dimensions we obtain with this choice the Lagrangian 
with canonically normalized Einstein-Hilbert term
\beq
- \frac{V^{(0)}_{s}}{2 \kappa^2} \int_{M^{5}} \sqrt{g^{(5)}} R^{(5)}
- \frac{V^{(0)}_{s}}{ 8 \pi (4 \pi \kappa^{2})^{2/3}}  \int_{M^4}
 \sqrt{g^{(4)}}
e^{3 \beta} F^2
\eeq
The interesting observation concerning this Lagrangian  
is that any further Weyl rotation of the 5d metric leaves the kinetic term 
of the gauge fields invariant. Hence, in principle, one could use some other frame in five dimensions, for instance the brane-frame. However,
in this paper we stay within the spirit of \cite{elpp} and choose 
5d canonical metric. Then, to make the connection of the parameters of the 
Lagrangian with the four dimensional Planck scale one should reduce the 
gravitational action further to four dimensions. To this end let us define 
$g^{(5)}_{5 \, 5} = e^{2 \gamma}$. Then the gravitational action in 
four dimensions, in terms of the metric which is canonical in 5d,
 takes the form
\beq
- \frac{2 V^{(0)}_{s} \pi \rho_0}{2 \kappa^2} \int_{M^{4}} \sqrt{g^{(4)}}
e^{\gamma} R^{(4)}
\eeq
where $2 \pi \rho_0$ is the coordinate length of $S^1$. The physical 
distance between the walls, measured with respect to canonical 
five dimensional metric, is $\pi \rho_p = e^{\gamma} \pi \rho_0$. 
Similarly as in the weak string case we have here three physical parameters,
which are $\kappa, \; \gamma, \; \beta\,$, and two reference numbers:
$V^{(0)}_s , \; \rho_0\,$. The first of the two numbers, the fiducial volume,
we choose in such a way that $e^{3 \beta} = 1/ g^{2}_{GUT}$, i.e.
$V^{(0)}_s = 2 \pi ( 4 \pi \kappa^{2})^{2/3}$. At this point one defines 
the four dimensional modulus $S$ through the requirement analogous to 
that employed in the reduction of the weakly coupled string, namely 
that it becomes the dynamical gauge coupling, $S_r = 1/g^2 = e^{3 \beta}$.  
The $\rho_0$ we choose in a 
more sophisticated way. We shall keep in mind that eventually we shall need to 
compare our results to the weakly coupled case. In the heterotic
string theory in ten dimensions there is a single dimensionful parameter, 
which is the string tension, or its inverse called $\alpha'$. Hence, 
it makes sense 
to measure all distances and mass scales in units which are suitable powers 
of the fundamental distance $\sqrt{\alpha'}$. Actually, this is the case 
in the reduction of the weakly coupled string action described earlier. There,
in the stringy frame,  $\sqrt{\alpha'}= \sqrt{2}\, \kappa_{10}/\lambda_{10}
= 4 \, \sqrt{\pi} / M_{Pl}$, and 
indeed, we have defined all the scales in terms 
of the Planck scale in that case. 
The correspondence between present $M$-theory 
parameters and the ten dimensional string parameters is easily 
obtained when one starts 
from the usual eleven dimensional Lagrangian and compactifies first down to 
ten dimensions. The resulting expression for $\alpha'$, again - in the string frame, is
\beq\label{e:corr}
\alpha' = \frac{1}{2 (4 \pi )^{2/3} \pi^2 } \frac{\kappa^{2/3}}{\rho_0}
\eeq
Since $\rho_0$ is the fundamental length scale of our model, then in order to 
be in agreement with the weakly coupled case normalization we shall 
take $\rho_0 = \sqrt{\alpha'}$ which gives
\beq \label{e:roo}
\rho_0 = \left ( \frac{\kappa}{4 \pi } \right )^{2/9} \frac{1}{2^{1/3} 
\pi^{2/3}}
\eeq
When one goes down to four dimensions, then exactly as in the effective Lagrangian for the weakly coupled case one tries to cast the supersymmetric 
part of the Lagrangian in the superfield language, and to this end defines 
two superfields whose kinetic terms do not mix (as in the weak case): 
$S_{s\, r} = e^{3 \beta}$ and $T_{s \, r} = e^{\gamma}$. 
It is interesting to realize that, with the normalizations we have assumed 
above, there exists a simple correspondence between weak and strong $S$ and 
$T$ 
fields. The simplest way to find it out is to start with the canonical 
metric in eleven dimensional $M$-theory Lagrangian and reduce it down 
to four dimensions in two ways: one way is first to go to  the canonical 
ten dimensional metric and then to 4d with the redefinitions of the 
metric chosen exactly as in the weak case, and the second way is to reduce the same 11d metric as in the strong case, first to 5d and then down to 4d.
Then, remembering that one has started from the same canonical 11d metric, one 
can compare the same entries of the original metric in both final 
parametrizations - strong and weak ones. Comparing both forms of 
$g^{(11)}_{11 \, 11}$
one obtains $\phi = e^{\gamma - \beta}$, and comparing the two expressions for 
$g^{(11)}_{M \, N }$ we obtain $e^{\beta} = \phi^{-1/4} e^{\sigma}$.
The net conclusion one can draw from these manipulations is that one can 
identify $S_s = S_w$ and $T_s = T_w$. It is interesting to note the surprising
fact, that the weak $S$ has been chosen in such a way, that it corresponds to 
the Calabi-Yau physical volume measured in `strong' eleven dimensional metric. 
Further to that, we can see that the `weak' and `strong' volumes of the same 
Calabi-Yau space are different. The fact that in each case we define the 
same physical $M_{GUT}^{6}$ to be equal to the inverse of the respective 
volume, which is not the same in `weak' and `strong' cases, amounts to the 
different relations between three fundamental physical parameters in 
both scenarios. To see what are the consequences of that difference, let us perform the fit 
to observables in the strongly coupled scenario. 
Now the fundamental relations are:
\beq \label{bas}
e^{3 \beta} = S_{sr} =\frac{1}{g^{2}_{GUT}}, \; M_{GUT}^{-6} = \frac{1}{g^{2}_{GUT}}
2 \pi (4 \pi \kappa^2)^{2/3}, \; \frac{M^{2}_{Pl}}{16 \pi } = \frac{ g^{2}_{GUT}}{   
M^{6}_{GUT} } \frac{\pi \rho_0 T_{sr}}{ \kappa^2}
\eeq
Substituting the same numbers  as before for four dimensional observables 
we obtain the physical distance 
between the hyperplanes or equivalently the mass scale 
at which the fifth dimension opens up $m_5 = (\pi \rho_p )^{-1} = 
0.8 \times 10^{16} \, GeV$, and the values of the moduli $S_{sr}=2.04$ and 
$T_{sr}=\frac{1}{\pi^{10/3}} \frac{1}{2^{19/3}}
( \frac{M_{Pl}}{M_{GUT}})^2 \frac{1}{S_{sr}^{1/3}} = 80 \;$. This allows us to compute the dimensionless 
string coupling  $g^{-2}_{s}= S_{sr}/T_{sr}^{3} = 4 \times 10^{-6}$. The inverse of 
this number is still large (although much smaller than the one obtained from the fit to the weakly coupled case),
but this time it is consistent with the initial assumption, which was that 
the underlying heterotic string {\em is} strongly coupled. 
To have more clear picture of the resulting pattern of scales 
let us quote the numerical values of the resulting eleven dimensional 
Planck scale $m_{11}= \kappa^{-2/9}$ 
and the ten dimensional string tension: $m_{11}= 0.2 \, / \sqrt{\alpha'}
= 4 \times 10^{16}\, GeV$.

It is interesting to discuss in the present context 
the definition  of the 
dimensionless gravitational coupling obtained from the stringy frame and    
check how close is the numerical unification of gauge and  dimensionless
gravitational coupling in the strongly coupled scenario. 
First, let us note that among the physical scales the scale of the fifth
dimension,
 $m_5$, is the lowest, and about five times {\em smaller} than the 
unification scale $M_{GUT}$. Then the stringy scale $1/ \sqrt{\alpha'}$ is 
about two times $M_{GUT}$, and the 4d Planck scale, which now plays 
the role of a low energy effective parameter, is the largest one. 
This means, that, as we have assumed, the unification with gravity 
takes place after the fifth dimension opens up. If we neglect, as we shall do here, the threshold effects from heavy modes on the circle and on Calabi-Yau 
space,
which are discussed in a forthcoming section, then the three gauge couplings run logarithmically without seeing the fifth dimension up to $M_{GUT}$,
but the dimensionless gravitational coupling changes its power-law running 
at $E=m_5$ from $\propto \, E^2$ to $\propto \, E^3$ (where $E$ 
is the energy scale). Just beyond  $M_{GUT}  $, the C-Y dimensions
open up and so the dimensionless  gauge coupling   $ {\tilde g}^2 $ 
scales like $\propto \, E^6 $ whilst  the dimensionless gravitational
coupling   scales like $ E^9 $ (recall that 
in a space time with more than four dimensions the gauge coupling is no longer dimensionless, and it scales with energy. In $d$ dimensions the dimensionless gauge coupling is  $\tilde{g}^2 = g^2 \, E^{d-4}$, and the 
dimensionless gravitational coupling is 
$g^{2}_{grav} = 16 \pi G^{(d)}_N \, E^{d-2}$. )

 To estimate the value of the numerical unification scale,  $E^{(M)}$, 
we note that in terms of the low energy physical scales the dimensionless couplings at energies higher than $M_{GUT}$ are
\beq \label{muni}
\tilde{g}^2 = \frac{g^{2}_{GUT} E^6}{M^{6}_{GUT}}, \; \; 
g^{2}_{grav} = \frac{16 \pi G_N}{m_5 M^{6}_{GUT}} E^9
\eeq
A quick computation shows that these couplings meet at $E^{(M)}\approx 1/\sqrt{\alpha'}$. This scale is essentially the string scale and 
lies in the region where the field theoretical description cannot be trusted.
However, the result of this simple calculation shows that in a more complete 
formulation of $M$-theory the idea of grand unification, in terms of the 
numerical unification of the couplings we defined above,  can work.  

It is interesting to note, that if one performs such a naive dimensional 
running in the weakly coupled string case, this time taking $d=10$ both in gauge and in the gravity sectors, one obtains numerical unification 
at $1/\sqrt{\alpha'}$. However, the common dimensionless coupling 
one obtains this way is orders of magnitude larger than one, signalling 
inconsistency of this approach in the weakly coupled case. 

\subsection{G as a Solution of the  Bianchi Identity}

As pointed out in \cite{wh} 
one can solve the Bianchi identity (\ref{e:5}) by defining  a modified 
  field strength \cite{strw,wh} 
\beq \label{e:6} 
G_{11ABC} = ( \partial_{11} C_{ABC} \pm 23\;  {\rm terms } + 
\frac{\kaps}{\sqrt{2} 
 \lambda^2} \sum_{m=1}^2 \delta^{(m)} (x^{11}) ( \omega^{(m)}_{ABC} - 
 \frac{1}{2} \omega^{(L)}_{ABC} ) 
 \eeq 
 where $\omega^{(m)} $, and  $ \omega^{(L)} $ are 
 ($E_8 )$ Yang Mills and Lorentz Chern-Simons 3 forms  defined on the 
respective 
boundaries.

Another way of solving Bianchi identity (\ref{e:5}), which does not lead 
to sigularities in the field strength $G_{IJKL}$ and in the equations of 
motion, is to assign the 
discontinuous limiting behaviour to the field strength components 
$G_{ABCD}$ when they pass across the boundaries:
\beq 
\lim_{x^{11}\rightarrow 0} G_{ABCD} = -\frac{3 {\kaps}}{
 \sqrt{2}   \lambda^2} 
 \theta (x^{11})\, ( {\rm tr}F^{1}_{[AB}\,F^{1}_{CD]} 
 - \frac{1}{2} {\rm tr} R_{[AB} \, R_{CD]} ), \;
\label{e:7} 
\eeq 
\beq 
\lim_{x^{11}\rightarrow \pi \rho_0} G_{ABCD} = -\frac{3 {\kaps}}{
\sqrt{2}
\lambda^2} 
 \theta (x^{11}- \pi \rho_0)\, ( {\rm tr}F^{2}_{[AB}\,F^{2}_{CD]} 
 - \frac{1}{2} {\rm tr} R_{[AB} \, R_{CD]} )
\label{e:8}
\eeq 
With this choice, and with definite $Z_2$-parity properties of all 
the bosonic fields, one can work on the half-circle $(0,\pi \rho_0)$
imposing the boundary conditions (\ref{e:7}),(\ref{e:8}),
instead of working on the full circle with singular configurations of $G$.
As pointed out in \cite{low10} algebraic manipulations 
allow then to convert Bianchi identity into eleven dimensional 
Laplace equation which has to be solved with boundary conditions (sources) 
following 
from (\ref{e:7}),(\ref{e:8}). It is not difficult to find out solutions 
for the components of the antisymmetric tensor field and its strength 
in terms of the sources located at the two fixed points. These solutions 
in turn determine the deviations from the underlying Calabi-Yau metric
required to maintain the $N=2$ supersymmetry in the five-dimensional bulk 
and $N=1$ chiral supersymmetry at the fixed hyperplanes. They can be read 
off from the formulae given in \cite{strw,low10}. It is possible,
using a momentum expansion  \cite{low10,llo}, to give explicit 
dependence of these solutions on the eleventh coordinate $x^{11}$.
The easiest way to go is to define the form $\Sigma$ which is Hodge dual 
to the field strength $G$, $\Sigma = d\Xi = \star G$, where $\Xi$ is the 
potential for $\Sigma$. Substituting it into the Bianchi identity, and 
supplying with the Lorentz-gauge condition $d^{\dagger} \Xi=0$, where 
$d^{\dagger}$ is the Hermitian conjugate of $d$, leads to the equation:
\beq \label{e:9}
\Delta_{11} \Xi_{J_1 J_2 J_3 J_4 J_5 J_6} = 0
\eeq
with boundary conditions  
\beq \label{e:10}
\lim_{x^{11} \rightarrow 0} \pa_{11} \Xi_{J_1 J_2 J_3 J_4 J_5 J_6} = 
\frac{1}{4 \sqrt{2} \pi} \left ( \frac{\kappa }{4 \pi } \right )^{2/3} 
(\star I^{1})_{J_1 J_2 J_3 J_4 J_5 J_6}
\eeq
\beq \label{e:11}
\lim_{x^{11} \rightarrow \pi \rho_0} \pa_{11} \Xi_{J_1 J_2 J_3 J_4 J_5 J_6} = 
 \frac{1}{4 \sqrt{2} \pi} \left ( \frac{\kappa }{4 \pi } \right )^{2/3} 
(\star I^{2})_{J_1 J_2 J_3 J_4 J_5 J_6}
\eeq
where 
\beq \label{e:12}
(\star I^{i})_{J_1 J_2 J_3 J_4 J_5 J_6} = \frac{\sqrt{g}}{6!} 
\epsilon_{J_1 J_2 J_3 J_4 J_5 J_6 J_7 J_8 J_9 J_{10}} 30 
(tr (F^{(i)[J_7 J_8 } F^{(i) J_9 J_{10}] })- \frac{1}{2} 
 tr (R^{[J_7 J_8 } R^{ J_9 J_{10}] }))
\eeq
and $i=1,2$ counts the fixed hyperplanes.
In specific cases these formulae simplify, for instance 
when one chooses all indices on $F$ and $R$ to be Calabi-Yau 
indices, one can factorize out the antisymmetric tensor of the 
visible four dimensions, and through the ansatz $ \Xi_{\mu_1 \mu_2
\mu_3 \mu_4 J_5 J_6} = \epsilon_{ \mu_1 \mu_2
\mu_3 \mu_4} \Xi_{J_5 J_6} $ one obtains a simplified 
problem for the $(1,1)$ 2-form $\Xi_{a \bar{b}}$. We shall return to this 
case in a moment. The other example consists in taking all indices 
on $F$ and $R$ to be noncompact indices. Then 
$\star I^{i} = \frac{1}{5!} \epsilon_{ \mu_1 \mu_2
\mu_3 \mu_4} 30 
(tr (F^{(i)[\mu_1 \mu_2 } F^{(i) \mu_3 \mu_4] })- \frac{1}{2} 
 tr (R^{[\mu_1 \mu_2 } R^{ \mu_3 \mu_4] })) V_6$
where $V_6$ is the  six dimensional volume form. After 
defining $\Xi_{J_1 J_2 J_3 J_4 J_5 J_6} = \frac{1}{5!} \Xi V_6$
one obtains  equation for the function $\Xi$. 
Similarly, one can solve for the antisymmetric tensor field in cases when 
sources with mixed, noncompact and compact, indices are excited. 
For the purpose of the present Section we shall be dealing with the pure cases 
given above. 
The solution for the antisymmetric field strength with purely non-compact 
indices is
\beqa \label{e:13} 
G_{\mu \nu \rho \delta} &=& - \frac{3 }{ 2 \sqrt{2} \pi } \left ( \frac{\kappa}
{4 \pi} \right )^{2/3} \left ( (tr (F^{(1)}_{[\mu \nu} F^{(1)}_{ 
\rho \delta] })- \frac{1}{2} 
 tr (R_{[\mu \nu } R_{ \rho \delta] })) (1 - \frac{x^{11}}{\pi \rho_0}) \right .
\nonumber \\ 
&-& \left . \frac{x^{11}}{\pi \rho_0} (tr (F^{(2)}_{[\mu \nu} F^{(2)}_{ 
\rho \delta] })- \frac{1}{2} 
 tr (R_{[\mu \nu } R_{ \rho \delta] })) \right )
\eeqa
and the solution for the antisymmetric field strength with purely compact 
indices is obtained by the replacement of the indices $\mu,\, \nu, \, 
\rho, \,  \delta$ by the suitable Calabi-Yau indices $I_1, \, I_2, \, I_3, \,
 I_4$.
In fact, the 
solutions with mixed indices assume in the approximation we use here 
exactly the same form with corresponding indices on the sources. 
The result (\ref{e:13}) implies through the relation $G=dC$ the form of the background part of 
$C_{JKL}$: 
\beq \label{e:15}
C_{JKL} =  - \frac{1 }{ 4 \sqrt{2} \pi } \left ( \frac{\kappa}
{4 \pi} \right )^{2/3} \left ( (\omega_{3}^{(1)YM} - \frac{1}{2} 
 \omega_{3}^{L}) (1 - \frac{x^{11}}{\pi \rho_0}) 
- \frac{x^{11}}{\pi \rho_0} (\omega_{3}^{(2)YM} - \frac{1}{2} 
\omega_{3}^{L} ) \right )
\eeq
However, this also implies through the 
relation $G_{11JKL} = (\,\pa_{11} C_{JKL}
\pm 23 \, perm\,)$ the specific form of $G_{11JKL}$:
\beq \label{e:16} 
G_{11JKL}= \frac{1} {4 \sqrt{2} \pi^2 \rho_0} \left ( \frac{\kappa}{4 \pi } \right )^{2/3} \left ( \omega_{3}^{(1)YM} +\omega_{3}^{(2)YM} - \omega_{3}^{L}
\right)
\eeq
We want to point out that 
this is an interesting expression, as it looks gauge non-invariant since  the
Yang-Mills Chern-Simons form is noninvariant under the gauge transformation 
$\delta A_I = - D_I \epsilon$ with $\delta \omega_{3\; IJK}^{(i) \, YM}= 
\pa_I (tr \epsilon F_{JK}^{(i)})$. To assure the gauge invariance of $G_{11JKL}$ one 
needs to require that the `free' part of the three-form $C_{IJK}$ 
is gauge non-invariant and transforms as 
\beq \label{e:17}
\delta C_{11IJ} = - \frac{1}{18 \sqrt{2} \pi^2 \rho_0 } \left ( \frac{\kappa}
{4 \pi} \right )^{2/3}  tr(\epsilon F_{IJ})
\eeq
It is an easy exercise to substitute this variation into the 
eleven-dimensional integral of the $C \wedge G \wedge G$ together with 
solutions for the $G_{MNPQ}$ and to perform the integration over $x^{11}$
in the resulting expression. This way one obtains the term which 
exactly cancels the usual gauge anomaly coming from ten-dimensional 
majorana-weyl gauginos. We discuss this point to show the slight difference 
in the ways the anomaly cancellation takes place depending on the way the Bianchi identity is solved. In the classic case discussed in \cite{wh} 
the required gauge variation is supported only on respective boundary, 
whereas here $C$ is uniformly transformed in the whole 
space, and variation becomes 
ten-dimensional after integrating over the explicit dependence of 
the integrand on $x^{11}$ obtained through solution of equations of motion.  

Finally, to illustrate more explicitly the way non-zero modes of the 
six dimensional Laplacian enter solutions of the Bianchi identity, 
let us consider the first example mentioned below the formula (\ref{e:12}).  
Let us assume that the sources on the fixed planes have the decomposition 
in the eigenfunctions of that Laplacian of the form 
\beqa \label{e:break1}  
\sum_i h^{(1) \,i} \pi^{i}_{I J} &=& - \frac{3 }{ 2 \sqrt{2} \pi } \left 
( \frac{\kappa}
{4 \pi} \right )^{2/3} \epsilon_{I J I_1 I_2 I_3 I_4} \left ( tr (F^{(1) \, [I_1 I_2} F^{(1)\, 
I_3 I_4 ] })- \frac{1}{2} 
 tr (R^{[I_1 I_2 } R^{ I_3 I_4 ] }) \right )
\nonumber  \\
\sum_i h^{(2) \,i} \pi^{i}_{I J} &=& - \frac{3 }{ 2 \sqrt{2} \pi } \left 
( \frac{\kappa}
{4 \pi} \right )^{2/3} \epsilon_{I J I_1 I_2 I_3 I_4} \left ( tr (F^{(2) \, [I_1 I_2} F^{(2)\, 
I_3 I_4 ] })- \frac{1}{2} 
 tr (R^{[I_1 I_2 } R^{ I_3 I_4 ] }) \right ) \nonumber \\
&& 
\eeqa 
In what follows, when we shall distinguish zero modes among $(1,1)$ forms 
$\pi^i$ then we shall call these zero modes $\omega^i$.
In terms of the mode expansion the solution of the relevant version of the 
equation (\ref{e:7}),(\ref{e:8}) with boundary conditions (\ref{e:10}), (\ref{e:11})
is:
\beqa \label{e:break2}
\Xi_{a \bar{b}}&=& \sum_{heavy \; \; modes} \pi^{i}_{a \bar{b}} 
\left ( - \frac{1}{48 \pi \rho_0} \frac{1}{\lambda_{i}^{2} } ( h_{(2) \, i} +
 h_{(1) \, i} )
+ \frac{1}{48} \frac{1}{\lambda_i } ( h_{(2) \, i} +
 h_{(1) \, i} ) \right .
\frac{\cosh ( \lambda_i x^{11} )}{
\sinh ( \pi \rho_0 \lambda_i ) } \nonumber \\
&-&\left . \frac{1}{48} \frac{ h_{(1) \, i}}{\lambda_i} \frac{\sinh (x^{11} \lambda_i
- \frac{\pi \rho_0 \lambda_i }{2} )}{ \cosh ( \frac{\pi \rho_0 \lambda_i }{2})}
 \right )  \\
&-&\sum_{zero\; \; modes} \omega^{i}_{a \bar{b}} \left (
 - \frac{1}{48 \pi \rho_0 } ( h^{(0)}_{(2) \, i} +
 h^{(0)}_{(1) \, i} ) ( \frac{(x^{11})^2 }{2} - \frac{\pi^2 \rho_0^2}{3 !} )
+ \frac{1}{48} h^{(0)}_{(1) \, i} ( x^{11}- \frac{\pi \rho_0 }{2} ) \right ) + \chi
\nonumber 
\eeqa
where $\chi$ is the solution of the equation $\Delta_X \, \chi_{a \bar{b}}
 = - 1/( 48 \pi \rho_0 ) \, \pi^{i}_{a \bar{b}} ( h_{(2) \, i} +
 h_{(1) \, i} )$ with trivial boundary conditions. The eigenvalues $\lambda_i$
are defined through the relation $\Delta_X  \, \pi^{i}_{a \bar{b}}=
 - \lambda_{i}^{2} \,  \pi^{i}_{a \bar{b}}$. It is easy to see that with the 
above definitions we have $G_{IJKL} = - \epsilon_{IJKLMN} \pa_{11} \Xi_{MN}$. 
We note, that at the order linear in $x^{11}$, which is the order to which 
we solve in this paper the Horava-Witten model, the antisymmetric tensor field 
strength obtained from (\ref{e:break2}) gives (\ref{e:13}).
The $(1,1)$ form $\Xi_{a \bar{b}}$ which  we have found here 
in the case of a general embedding is also interesting, as it gives directly 
corrections to the metric of the Calabi-Yau space, see \cite{low10}. 
One of the important findings of the next Section is that for a general 
embedding the coefficients of the zero modes in the decompositions of sources 
fulfil the equality
\beq \label{e:h0}
 h^{(0)}_{(2) \, i} +
 h^{(0)}_{(1) \, i} = 0
\eeq
which is due to the requirement that the integrability condition for the 
Bianchi identity be fulfilled. The obvious consequence is that the solution
given above simplifies in its zero-mode part, and that the zero-mode 
component of $G_{MNPQ}$ (with all indices along C-Y space) is independent of 
$x^{11}$ in the general case. 

\section{Axionic Thresholds}
In this section we shall see that the solution of the Bianchi identity
in the previous section, for general embeddings, implies that
there will be axionic threshold corrections in the $d = 4 $ theory.   
The condition that the Bianchi identity has a solution is that the 
source terms on its right-hand-side add up to zero in cohomology, i.e. 
that the cohomology class $ [ \sum_{i=1,2} trF^{(i)} \wedge  
F^{(i)} - tr R \wedge R ] $ 
is trivial. 
The discussion given above does not depend thus far  in any way on the 
particular embedding one uses to solve this integrability condition. 
The standard embedding consists in identifying  the spin connection of the 
compact six dimensional space with  the 
$SU(3)$ subgroup of one of the two $E_8$ gauge groups present in the theory,
$F^{(1)}_{IJ}=R_{IJ}$. 
This leads to the model which has one of the $E_8$ factors broken down to 
$E_6$, with chiral matter in four dimensions. 
This is the solution which has been discussed so far 
in the context of the Horava-Witten model. 
However, this is by no means the unique possibility and, in fact, other 
embeddings, called non-standard,  give models which have the 
group theoretical structure even more appealing than the simple $E_6$.
Hence it is interesting to see whether the basic features of the naive 
four-dimensional limit of the Horava-Witten model change in any significant way
when one departures from the standard embedding. 
As pointed out in \cite{elpp,low5} the fundamental low energy theory 
coming from the compactification of that model on the deformed Calabi-Yau
space lives in five dimensions. However, at least for the gauge sector 
which is basically four-dimensional after compactification, 
it makes sense to define the effective nonrenormalizable
 four-dimensional action which 
is globally supersymmetric at the lowest nontrivial order. We are primarily 
interested in the effective gauge kinetic terms. 
These terms, assuming four-dimensional supersymmetry, are
\beq \label{e:18}
L_{gauge} = - \frac{1}{4} \int d^4 x  \; tr(F_{\mu \nu} F^{\mu \nu})
\, \Re (f) 
+ \frac{1}{8} \int d^4 x \; \epsilon^{\mu \nu \rho \delta} \,
 tr(F_{\mu \nu} F_{\rho \delta}) \, \Im (f)
\eeq
We are interested in
the form of the function $f$ in the visible and hidden sectors. In terms 
of the moduli fields it 
should follow from the Kaluza-Klein reduction of the eleven-dimensional 
theory, or can be read off the effective Lagrangian of the weakly coupled 
heterotic string, as argued in \cite{bd}, \cite{sstieb}. 
It is the first path that we shall follow, although at the end we turn back to 
the correspondence with the weakly coupled result. 
The easiest way to identify the particular combination of moduli 
which forms gauge kinetic functions of gauge groups on each of the 
fixed hyperplanes is to find out all the contributions to axionic couplings,
i.e. the coefficients multiplying the four dimensional operators 
$\epsilon^{\mu \nu \rho \delta} tr(F_{\mu \nu}^{(i)} F_{\rho \delta}^{(i)})$.
One obvious contribution is the one coming from the bulk kinetic terms for the 
antisymmetric tensor field enhanced by the Chern-Simons forms implied by the 
sources in the Bianchi identity. However, it is easy to convince oneself that 
the massless axion which is the trivial zero-mode of $C_{11 \mu \nu}$ 
couples in exactly the same way to both $\epsilon^{\mu \nu \rho \delta}
 tr(F_{\mu \nu}^{(i)} F_{\rho \delta}^{(i)})$, $i=1,2$.
Since the difference between the axionic couplings coming from the 
kinetic terms vanishes, we have to look for an additional source
of axionic couplings which would give a nonzero difference which 
shall partner  the difference of the Calabi-Yau 
volumes at both planes (the difference of the volumes has been computed 
in the embedding-independent way in \cite{strw}). Such terms are easily 
found  through the 
Kaluza-Klein compactification of the eleven 
dimensional $C \wedge G \wedge G$ term. The relevant integral is that of 
the last term in  formula (\ref{e:1}). The part which gives axionic 
thresholds must be proportional to $G_{\mu \nu \rho \delta} G_{IJKL} 
C_{11MN}$. The components $C_{11MN}$ have an expansion in terms of the 
eigenmodes of the six-dimensional Laplacian $\Delta_X$. The harmonic forms 
which are zero-modes shall be denoted by $\omega^{\alpha}_{IJ}$ and the nonzero
modes by $\pi^{\alpha}_{IJ}$ with $\Delta_X \pi^{\alpha}_{IJ} 
= - (\lambda^{\alpha})^2 \pi^{\alpha}_{IJ}$. Then we expand 
\beq \label{e:21}
C_{11MN} = \frac{2 \sqrt{2}}{3} ( \theta^{\alpha}(x) \omega^{\alpha}_{MN} +  
\bar{\theta}^{\beta}(x) \pi^{\beta}_{MN} )
\eeq
where $x$ denotes four noncompact coordinates plus $x^{11}$. 
One should note that we should also expand the $\theta$s with respect to 
$x^{11}$. The lowest order terms are (all functions are periodic 
with the period $2 \pi \rho_0$ and $\theta$s are $Z_2$-even):
\beq \label{e:22}
\theta^{\alpha}(x^{\mu};x^{11})= \theta^{\alpha}(x^{\mu}) +
\theta^{\alpha}_1 (x^{\mu}) \frac{1}{\sqrt{\pi \rho_0}}
 \cos (\frac{x^{11}}{\rho_0}) + \ldots
\eeq
(and similarly for barred thetas).
Let us consider the massless axions $\theta^{\alpha}(x^{\mu})$.  
We substitute into the last term of eq.(1) the explicit solutions
(\ref{e:13}), with the linear dependence on the $x^{11}$.
In principle we should perform the integrals over Calabi-Yau first,
leaving the five-dimensional coupling between the gauge fields and 
massless axions with the explicit $x^{11}$ dependence. However, 
we are interested in four-dimensional expressions, hence we can 
integrate over the circle. After simple algebraic manipulations 
we obtain:
\beqa \label{e:23}
\delta  L^{(4)} &&= 
 \frac{1}{ \kappa^2} \frac{ \rho_0}{24 \pi} \left (
\frac{\kappa}{4 \pi} \right )^{4/3}  tr(F^{(1)} \tilde{F}^{(1)})
\left [ \theta^{\alpha}  \int_X \omega^{\alpha} \wedge ( tr(F^{(1)} 
\wedge F^{(1)})
- \frac{1}{2} tr(F^{(2)} \wedge F^{(2)}) \right .\nonumber \\
& -&  \left . \frac{1}{4} tr(R
  \wedge R ) ) 
+ \bar{\theta}^{\beta}  \int_X \pi^{\beta} \wedge ( tr(F^{(1)} \wedge F^{(1)})
- \frac{1}{2} tr(F^{(2)} \wedge F^{(2)}) - \frac{1}{4} tr(R \wedge R ) ) \right ]
\nonumber \\
&+& \frac{1}{ \kappa^2} \frac{ \rho_0}{24 \pi} \left (
\frac{\kappa}{4 \pi} \right )^{4/3} tr(F^{(2)} \tilde{F}^{(2)})
\left [ \theta^{\alpha}  \int_X \omega^{\alpha} \wedge ( tr(F^{(2)} \wedge F^{(2)})
- \frac{1}{2} tr(F^{(1)} \wedge F^{(1)}) \right . \nonumber \\
 &-& \left . \frac{1}{4} tr(R \wedge R )) 
+ 
\bar{\theta}^{\beta}  \int_X \pi^{\beta} \wedge ( tr(F^{(2)} \wedge F^{(2)})
- \frac{1}{2} tr(F^{(1)} \wedge F^{(1)}) - \frac{1}{4} tr(R \wedge R  )) \right ]
 \nonumber \\
&&
\eeqa
This formula holds for any embedding, standard or not. 
The question arises about the correlations between specific 
integrals over Calabi-Yau space entering the formula above. First,
one has to note that in all these integrals one can take the same Calabi-Yau space, as taking the variation of C-Y would lead to corrections which are 
higher order in $\kappa$, beyond the order to which the 
Horava-Witten model is defined. Then one has to consider separately integrals 
with $\omega^\alpha$ and with $\pi^\beta$. Let us consider the integrals 
containing harmonic forms. If one adds integrals with the same form 
$\omega^\alpha$ which are coefficients of $F^{(1)} \tilde{F}^{(1)}$
and $F^{(2)} \tilde{F}^{(2)}$ respectively, lets call them 
$\gamma^{\alpha}_1$ and $\gamma^{\alpha}_2$, one gets 
\beq \label{e:230}
\gamma^{\alpha}_1 + \gamma^{\alpha}_2
= \frac{1}{ \kappa^2} \frac{ \rho_0}{24 \pi} \left (
\frac{\kappa}{4 \pi} \right )^{4/3}
\frac{1}{2} \int_X \omega^{\alpha} \wedge \left ( tr( F^{(1)} \wedge F^{(1)})+
tr(F^{(2)} \wedge F^{(2)}) - tr(R \wedge R \right ) )
\eeq
and we remeber that $[ F^{(1)} \wedge F^{(1)}+
F^{(2)} \wedge F^{(2)} - R \wedge R] \, = \, 0$. If one restricts oneself
to compact indices on the forms, then the reasoning from the weak 
string case can be adopted and one can write $dH= ( tr(F^{(1)} \wedge F^{(1)})+
tr(F^{(2)} \wedge F^{(2)}) - tr(R \wedge R) )$. The integral of $dH$ over 
any closed four-dimensional surface must vanish, as $H$ must be globally 
defined (it enters the energy density under the name of $G_{11IJK}$). 
If so, then for any embedding, the above integral is 
\beq \label{e:24}
\gamma^{\alpha}_1 + \gamma^{\alpha}_2
= - \frac{1}{ \kappa^2} \frac{ \rho_0}{24 \pi} \left (
\frac{\kappa}{4 \pi} \right )^{4/3}
\frac{1}{2} \int_X d \omega^{\alpha} \wedge H = 0
\eeq
because $\omega^{\alpha}$ is a harmonic form. 
In exactly analogous way one proves the relation (\ref{e:h0}).
Next, we also consider the contribution from the non-zero modes 
$\pi^{\beta}$.  
The sum of respective coefficients $\bar{\gamma}^{\beta}_1$ and $\bar{\gamma}^{\beta}_2$ is 
\beqa \label{e:25}
\bar{\gamma}^{\beta}_1 + \bar{\gamma}^{\beta}_2
&=& \frac{1}{ \kappa^2} \frac{ \rho_0}{24 \pi} \left (
\frac{\kappa}{4 \pi} \right )^{4/3}
\frac{1}{2} \int_X \pi^{\beta} \wedge \left (  tr(F^{(1)} \wedge F^{(1)})+
tr(F^{(2)} \wedge F^{(2)}) - tr(R \wedge R \right ) ) \nonumber \\
&=&-\frac{1}{ \kappa^2} \frac{ \rho_0}{24 \pi} \left (
\frac{\kappa}{4 \pi} \right )^{4/3}
\frac{1}{2} \int_X d \pi^{\alpha} \wedge H \neq 0
\eeqa
since in general $d\pi \neq 0$. 

It should be stressed, that the threshold corrections we have given here
are corrections to the {\em unified gauge couplings on the respective
walls}
 for any embedding. After breaking $E_8 \times E_8$ down to a group containing 
a subgroup $G$, one should in principle change the normalization of generators 
$T$  
of $G$, using the relation  
$Tr_{Adj( E_8 \times E_8 )} T^2 = f \, Tr_{Adj(G)} T^2$
where $f$ turns out to be a small number ($<\, 10\,$). Once the detailed decomposition of both $E_8$s is known, one can read-off 
corrections to the individual group factors.  
In the next Section we shall discuss a
particularly interesting class of embeddings which
mix the 
subgroups of the two $E_8$s. 

There are two important conclusion to be drawn from this reasoning. 
Firstly, if one restricts oneself to true four-dimensional  pseudoscalar 
zero modes, which we call here $\theta^\alpha$, then the above calculation 
gives us canonical  individual axionic threshold corrections 
to each separate gauge group. This is not so for the corrections to the 
coefficients of the $F^2$ terms, which come from the difference of the volumes of the C-Y spaces on both walls. There one computes directly the 
difference of the volumes, and in a general case we do not get any 
 canonical reference point,
which would tell us  how the difference should be split between the planes.  
\footnote{In the case of the standard embedding one can 
argue that such a point emerges from the calculation of the correction to the background metric, and lies in the middle of the interval between walls, where 
the correction vanishes.}.
Secondly, the canonically chosen axionic threshold corrections (and, assuming 
underlying $N=1$ 4d supersymmetry, full universal threshold corrections) 
are equal in magnitude, and of opposite sign if one restricts oneself to 
the zero-mode harmonic forms in the C-Y decomposition of the $C_{11AB}$. 
However, if one goes a step further, and considers the non-zero modes on 
Calabi-Yau, then the exact correlation of the threshold coefficients 
between walls is violated. This is actually not surprising, as the integration 
with the zero modes over C-Y space extracts only the averaged
part of the integrands, whereas the higher modes are sensitive to the
finer structure of these. 
Of course, the contributions of the higher modes are weighted by the 
expectation values of the four-dimensional fields $\bar{\theta}^\beta$. 
These fields have masses of the order of $V^{-1/6}_{CY}$, hence at low 
energies these expectation values are expected to be zero. 
However, when one reaches up in energies, considering for instance
the details of the unification of gauge and gravitational couplings 
in the model, then these modes should be switched on around the scale 
where the largest of the six C-Y dimensions peels off. There,
the expectation values should be put at $< \, \bar{\theta}^\beta \, > \, = 
\, 1/ R_{max} $ rather than at zero\footnote{Note that whenever necessary 
dimensions are supplied by suitable powers of $1/\sqrt{\alpha'}$.}.  

Of course, once we consider the higher modes of the Laplacian on the C-Y
space, we should consider also higher Fourier modes on the circle 
along the eleventh dimension. This is because one expects 
that the radius of the eleventh dimension is smaller than the largest 
radius of C-Y, $\pi \rho \, < \, R_{max}$. 
It is easy to compute the coefficients associated with the higher modes
 on the circle. The coefficients corresponding to the first nonhomogeneous 
modes $\theta_{1}^\alpha$ are:
\beqa \label{e:26}
\delta_{1}  L^{(4)} &=& \frac{1}{ \kappa^2} \frac{\sqrt{ \rho_0}}
{18 {\pi}^{7/2}} \left (
\frac{\kappa}{4 \pi} \right )^{4/3}  tr(F^{(1)} \tilde{F}^{(1)})
\left [ \theta_{1}^{\alpha}  \int_X \omega^{\alpha} \wedge ( tr(F^{(1)} 
\wedge F^{(1)}) -  \frac{1}{2} tr(R \wedge R) ) \right ] 
\nonumber \\
&-& \frac{1}{ \kappa^2} \frac{\sqrt{ \rho_0}}
{18 {\pi}^{7/2}} \left (
\frac{\kappa}{4 \pi} \right )^{4/3}  tr(F^{(2)} \tilde{F}^{(2)})
\left [ \theta_{1}^{\alpha}  \int_X \omega^{\alpha} \wedge ( tr(F^{(2)} 
\wedge F^{(2)}) - \frac{1}{2} tr(R \wedge R) ) \right ]
\eeqa
Hence, as one can easily check using the argument applied earlier 
in this section, at this level the respective 
threshold coefficients on the walls are equal and of the same sign
\beq \label{e:28} 
\gamma_{(1)\,  1}^{\alpha} - \gamma_{(1)\, 2}^{\alpha} = 0
\eeq
Again, higher modes on the circle are actually expected to be excited 
when one reaches the energy scale at which the eleventh dimension becomes
visible, $m_5 = 1/(\pi \rho)$.
 There the expected value of 
$\theta_{1}^\alpha$ might be better approximated by its {\em rms} 
value $m_5/\sqrt{2}$ than by zero. 

Identifying the axions $\theta_i$ as the imaginary parts of 
the $h_{(1,1)}$ moduli 
$T_i$ (which generalize the situation with the single general modulus $T$),
on the basis of the equations (\ref{e:23}) and
(\ref{e:24}) and  using the assumption of four dimensional supersymmetry
 and holomorphicity of 4d supersymmetric gauge couplings (\ref{e:18}),
we obtain the following result for the gauge couplings
at the unification 
scale : 
\beq \label{e:ps}
1/ g^{2}_{(k)} = Re(S + \epsilon_{i\, (k)}  T_i )
\eeq
($i=1,\ldots,h{(1,1)}, \; k=1,2\,$) where the chiral field $S$ remains to be defined and the individual coefficients 
$\epsilon_{i\, (k)}$ are related through 
\beq
\epsilon_{i\, (1)}=
- \epsilon_{i\, (2)} =\gamma_{i \,(1)}
\eeq
This result is valid for a general embedding, standard or not.
 This is due to the fact, that the integration 
over Calabi-Yau with all external four dimensional fields massless
picks out only the massless contribution from the solution for $G$,
which leads to the above property of $\epsilon_{i\,(1)}$ and 
 $\epsilon_{i\,(2)}$ independently 
of the nature of the embedding.
Using eq.(27), eq.(32) and eq.(33)
the explicit form of the coefficients $\epsilon_{i\, (k)}$ reads
\beqa \label{e:xpl}
\epsilon_{i\, (1)} &=& - \frac{ \pi \rho_0 }{2 (4 \pi)^{4/3} \kappa^{2/3}}
\frac{1}{8 \pi^2} \int_X \omega_{i} \, \wedge \, (\, trF^{(1)} \wedge F^{(1)}
 - \frac{1}{2} \, trR \wedge R \,) 
\eeqa
where $\omega_i$ are massless $(1,1)$-forms, and 
the topological integral over C-Y space shall be 
paramentrized in terms of instanton numbers in Section 5. 

The difference of the couplings given above 
is  of the relative order $o(\kappa^{4/3})$ 
with respect to the gravitational part of the Lagrangian, and represents  
the difference of volumes 
of the two C-Y spaces  on the hidden and visible walls. 
When we  
incorporate these corrections into the effective four dimensional model,
it constitutes a departure from the order $(\kappa^{2/3})$ situation 
discussed earlier in Section 2.1 . 
Hence,   
before we proceed with the discussion we need to make some comments 
on the interpretation
 of the `strong' fields $S$ and $T$ in the present setup. 
As pointed out in \cite{bd} in the correct five dimensional theory 
there exists a scalar field belonging to the universal hypermultiplet,
let us call it here ${\cal S}$, whose expectation value measures  the 
C-Y volume along the fifth dimension. When one tries to solve the 
equation of motion for this field along the fifth dimension, the 
step which is necessary to obtain four dimensional model, 
then one encounters a zero mode of this field, which is homogeneous in the 
direction of $x^5$. This is a good candidate for the effective 
4d $S$. In this paper we work in a simplified setup, where to get the 4d
model we solve only the 11d equations of motion, where one obtains a simple 
linear dependence of the volume on $x^5$. The zero mode of such a linearly 
changing quantity can be defined as the arithmetic mean of the limiting values on the boundaries. When we define the effective $S$ in such a way, as 
$S_r =(V(x^{11}=0) + V(x^{11}= \pi \rho_0))/ 2 V^{(0)}_s$ and choose 
the reference volume according to the normalization assumed in the
Lagrangian containing only terms of relative order $o(\kappa^{2/3})$, then,
indeed, in terms of  this  $S$ the holomorphic gauge
 couplings look like $S \pm \Delta(\kappa^{4/3})$.
Hence, what we described above is the five dimensional interpretation of 
the `strong' field $S$ we are going to use in the rest of the paper. 
As for the overall modulus $T$, the degree of freedom it contains 
does exist in 
five dimensions 
as the component  $e_{5 \, 5}$ of the vielbein and belongs to the 
5d gravitational 
multiplet (together with graviton and graviphoton, whose fifth component is 
the axial part of $T$). There is no trouble with defining the correct 
effective $T$: it is simply the $(\int_{0}^{\pi \rho_0} e_{5 \, 5})/ 
(\pi \rho_0) $. Analogous treatment applies to additional moduli $T_i$ 
which from five dimensional point of view are parts of $h_{(1,1)} -1$ vector 
supermultiplets. 
The definitions of $\rho_0, \; \alpha'$ and the 
relation between $T$ and four dimensional $M_{Pl}$ we   leave 
as defined in Section 2.1 since we do not want to go beyond the validity 
limit of the Horava-Witten construction. 
This summarizes the interpretation 
of $S$ and $T$ in the context of effective 4d Lagrangian containing 
 terms of relative order $o(\kappa^{4/3})$. 

\section{Embeddings that Mix Gauge Sectors from Different Walls and Anomalous $U(1) $}

The interesting aspect of non-standard embeddings in M-theory 
which deserves a separate discussion is the possibility of obtaining 
matter with charges that mix visible and hidden sector  gauge groups.
It should be stressed that this is the necessary condition
for the presence of an anomalous $U(1)_A$ group, with  
cancellation of the $U(1)_A$ anomaly through the 
gauge transformation of the universal axion, as it happens in 
stringy models with `anomalous' $U(1)_A$ factor. The standard embedding 
does not give mixed charges, hence does not give `anomalous' $U(1)_A$, 
neither in smooth C-Y compactifications, or in orbifolds compactifications. 
Before we discuss how it can exist in nonstandard embeddings, let 
us specify first what  we actually mean by nonstandard embeddings \cite{gsw,dist}. 

Roughly speaking, let us take a Calabi-Yau manifold, and a vector bundle 
which is a direct sum of stable holomorphic vector bundles,
$V=\bigoplus V_i$. Let us make the gauge fields of the gauge connections
on that bundle satisfy the Einstein-K\"ahler equations\footnote{Supplied with 
integrability conditions $\int_X \omega \wedge \omega \wedge F = 0$.}
\beq \label{e:29}
F_{a a}= F_{\bar{a} \bar{a}} = 0, \;\; \; \; g^{a \bar{a}} F_{a \bar{a}}=0
\eeq
Suppose that, in addition, the second chern classes of the gauge bundles $V_i$ 
sum up to the second chern class of the tangent bundle $T$,
\beq \label{e:30}
\sum c_2 (V_i) = c_2 (T)
\eeq
(which is equivalent to $[ \sum_i F^{(i)} \wedge F^{(i)}
- R \wedge R] \, = \,[ 0 ]$ encountered earlier), and integrability 
conditions for the equations of motion \cite{gsw}  are satisfied. 
After compactification to 
four dimensions, these conditions lead to unbroken $N=1$ supersymmetry. 
This abstract vacuum bundle is embedded into the 
actual $E_8 \, \times \, E_8$ gauge bundle, and $E_8 \, \times \, E_8$ 
is broken down to the subgroup $G$ formed by the generators which 
commute with the structure group $J$ of the vacuum bundle. 

If one takes $V_1 = T, \; \; V_2 =0$, where indices $1,\, 2$ 
correspond to the two $E_8$s, and one makes the natural identification 
of the $SU(3)$ holonomy group gauge fields with those of the $SU(3)$ 
subgroup of one of the $E_8$s, then all the equations are fulfilled, 
and the standard embedding is realized. However, it is well known by now 
that many other choices both for the vacuum bundle $V$ and for the 
actual embedding of the vacuum bundle into gauge bundle are possible,
in manifold compactifications and in orbifold compactifications 
for instance. One extension of the above construction to 
the Horava-Witten type 
supergravity is  when each item from the direct sum of vacuum bundles 
is embedded fully into one or the other $E_8$ bundle. 
However, the possibility which is in a sense most interesting is that 
a sub-bundle is  partially embedded into both $E_8$s. To be more 
specific let us take a line bundle with the structure group $U(1)$,
and the corresponding vacuum gauge field $(1,1)$ form ${\cal F}_{a \bar{a}}$.
Then let us take a subgroup $U(1)_1$ from one $E_8$ and a $U(1)_1$ 
from the other. Finally, let us define the embedding with the condition 
\beq \label{e:32}
 \left ( \cos (\theta) F_{(1) a \bar{a}} + \sin (\theta) 
F_{(2) a \bar{a}} \right ) 
=  {\cal F}_{a \bar{a}}
\eeq
where $F_{1 a \bar{a}}^{(1)} $ and $F_{2 a \bar{a}}^{(2)} $
are the corresponding $U(1)_1 \times U(1)_2 $ field strengths. The 
mixing angle is given by 
$\cos (\theta)=Tr_{Adj(E_8 \times E_8)} ( T^1 {\cal T})$
and $\sin (\theta)=Tr_{Adj(E_8 \times E_8)} ( T^2 {\cal T})$, with 
$T^1,\;T^2, \; {\cal T}$ being generators of the first and second $U(1)$ 
and of the structure group of the linear bundle.

Before we continue, it is worth pointing out that it is normal 
for the curvature of line bundles to lie in integral cohomology classes of
the C-Y manifold 
$X$. 
If this is the case with the bundles associated with $U(1)_1  $ and
$U(1)_2 $ then the question arises under what conditions can the 
curvature ${\cal F} $ in (\ref{e:32}) lie in integral cohomology classes?
Using the constraint that the orthogonal combination 
$F=-sin (\theta ) F_{1a \bar{a}}^{(1)} + cos(\theta ) F_{2 a \bar{a}}^{(2)} 
\, = \, 0 $, we see that we must take $sin (\theta ) \, = \, p/n \, , \, 
cos (\theta ) = q/n $ where $n^ 2 \, = \, p^2 + q^2 $, $p,q,n $ being
integers with $q,p $ labeling the cohomolgy classes of the $U(1)_1 $ and
$U(2)_2 $ bundles. If we substitute this back into the embedding
(\ref{e:32}) then it is clear that this equation now defines a new $U(1)$
bundle in a cohomology class labelled by the integer $n$, in terms of those
bundles whose classes are labelled by $q$ and $p$.

The kinetic terms for $F_1^{(1)} \, , F_2^{(2)} $ give  
\beqa \label{e:33}
\frac{1}{4 g_{1}^{2}} F_{1}^{2} &\rightarrow& 
\frac{1}{4 g_{1}^{2}} \left ( \cos^2 (\theta) {\cal F}^2
 - 2 sin(\theta ) cos (\theta) {\cal F} F + \sin^2 (\theta) F^2 \right )\cr  
&&\cr
\frac{1}{4 g_{2}^{2}} F_{2}^{2}
& \rightarrow & 
\frac{1}{4 g_{2}^{2}} \left ( \sin^2 (\theta) {\cal F}^2
+2 sin (\theta ) cos(\theta ) {\cal F}F+ \cos^2 (\theta) F^2 \right )
\eeqa
where in the above we have extended the definitions of 
$F$ and ${\cal F} $ to include noncompact indices.
The field 
${\cal F}$ is  `Higgsed' by one of the $h_{(1,1)}$ nonuniversal axions 
coming from the C-Y decomposition of the antisymmetric tensor field 
components $C_{11 a \bar{a}}$ and decouples from the massless spectrum 
\cite{ew}. 
This happens in the following way. Since the field strengths 
$F_{1a \bar{a}}^{(1)},  F_{2 a \bar{a}}^{(2)}$ define closed
$(1,1)$ forms on $X$ \footnote{We are considering 
manifolds with 
$h_{(1,1)}>1$.}, the  decomposition of  $C_{11 a \bar{a}}$
will include massless $d=4 $ axionic fields $a_1, a_2 $ , with 
$C_{11 a \bar{a}} \, = \,  a_1\,  F_{1a \bar{a}}^{(1)} + 
a_2 \, F_{2 a \bar{a}}^{(2)}$ . However from what we have said above,
 only the combination  $a_1 cos (\theta ) + a_2 sin (\theta)
$ in fact appears in this expansion. If we look at the  terms in the effective
$d=4$ action that contain $ C_{11 a \bar{a}} $, amongst these we find 
(using the form of the background part of $G_{11JKL} $ as given in 
(\ref{e:16}) )
$ \int_X d^6 y \sqrt{g^{(6)} } ( \partial_\mu  C_{11 a \bar{a}}  + \frac{1}{2 \sqrt{2} \pi^2 \rho_0 } 
(\frac{\kappa}{4 \pi} )^{2/3}  ( \, A^{(1)}_{1\,\mu } F_{1a \bar{a}}^{(1)}
+ A^{(2)}_{2\,\mu } F_{2 a \bar{a}}^{(2)} ) \,  )^2 $. This makes it clear, 
from the choice of our mixed embedding,  that   $ a_1 cos (\theta ) + a_2 sin (\theta) $
is eaten by gauge fields $ cos (\theta ) A^{(1)}_{1\,\mu }  +  sin (\theta) A^{(2)}_{2\,\mu } $  with field strength  
${\cal F}_{\mu \nu} $.

The orthogonal combination of  gauge fields with field strength  $F$ describes a massless $U(1)$ vector boson, that couples to charged particles on both walls. 
It can in principle correspond to an anomalous $U(1)_A$, as 
the trace of the associated generator over the massless spectrum does not 
have to be zero\footnote{Anomalous $U(1)_A$ in the Horava-Witten model
has been discussed recently 
in \cite{march} in different contexts.}. 
The puzzle comes from the fact that $U(1)$ gauge 
transformations on {\em different} and spatially separated 
hyperplanes are correlated. In fact, we have imposed the mixed embedding 
through the nonlocal  constraint 
equations (\ref{e:32}), which relates fields living on different fixed points
of the underlying orbifold in the direction of $x^{11}$. 
It is the Bianchi identity  
which resolves  the puzzle. First, let us remind the reader that the Bianchi 
identity holds not only at the level of vacuum configurations 
but, also, at the level of full fields, including fluctuations 
around vacuum. As we saw in Section 2. solving the Bianchi identity 
we obtain the configuration of the antisymmetric tensor field in the
bulk which is determined in non-perturbative way by 
the gauge fields on both  planes. 
The same mechanism works for our mixed $U(1)$. It is the field $G$ 
which provides communication through the bulk between the two 
`components' of the gauge field of this $U(1)$ . 
{}From equation (\ref{e:33}) one
obtains the effective four dimensional kinetic term for the mixed $U(1)$ 
of the form
\beq \label{e:35}
L_{F} = \frac{1}{4} \left ( \frac{\sin^2 (\theta)}{g_{1}^{2}} +  
\frac{\cos^2 (\theta)}{g_{2}^{2}} \right ) F^2
\eeq
Hence the effective four dimensional coupling is a  harmonic 
average of the two  couplings for groups coming exclusively 
from a single hyperplane:
\beq \label{e:36}
g_{mixed}^{2} = \frac{ g_{1}^2 g_{2}^2 }{\sin^2 (\theta) g_{1}^2
+ \cos^2 (\theta) g_{2}^2}
\eeq
Taking $1/g_{1}^2 \, = \, S_r +\sum_{i'} \epsilon_{i'} T_{r \, i'}$
and   $1/g_{2}^2 \, = \, S_r -\sum_{i'} \epsilon_{i'} T_{r \, i'}$
one obtains 
\beq \label{e:37}
\frac{1}{g_{mixed}^{2}} = S_r + \sum_{i'} \epsilon_{i'} T_{r i'} \, ( \sin^2 (\theta) 
- \cos^2 (\theta) )
\eeq
where $ i' \, = \, 1 ...h_{(1,1)} -1 $. We have excluded the axion
whose wave function on the C-Y space is proportional to 
${\cal F}_{a\bar{a}} $ as it is eaten in the `Higgsing' of 
${\cal F}_{\mu \nu} $.
One can check that the form of the moduli dependence in (\ref{e:37})
implies , via supersymmetry, axionic couplings to $F {\tilde F} $ in $d=4$
that can be explicitly verified by reduction of the $CGG$   term in the 
eleven dimensional action, using the explicit solutions to the Bianchi
identities for G.
The effective 
Lagrangian for the anomalous $U(1)$ realizing the M-theoretical version 
of the four dimensional Green-Schwarz mechanism 
shall have the form 
\beq\label{e:38}
L_{K} = - \log ( S + \bar{S} + c \, V_{U(1)_A} ) \left |_{D} \right . 
\eeq
with $c \propto 1/S_{r} = g^{2}_{mixed} \left |_{\mbox{{\rm lowest order in  
$\kappa$}}} \right . $, (see discussion at the end of this Section. )

The evaluation of   the anomaly coefficient $c$ 
can be achieved  in the following way. The form of the K\"ahler potential $L_K $ in 
(\ref{e:38}) arises as a consequence of the supersymmetrization of  the four 
dimensional terms  $ c \, \int d^4 x  B \wedge F $ , where $B$ is the  two form 
potential, whose dualized 3-form field strength  defines the pseudo-scalar field in $S_r $.
Such  a term is obtained from the reduction of the $CGG$ Chern-Simons term in the 
M-theory Lagrangian, when we substitute  the explicit solutions of the Bianchi
identities, eq. (\ref {e:13}),
 generalized to the case where we have mixed  
compact and non compact indices on  $G$, and we identify $C_{11 \mu \nu}
\,  = \, B_{\mu \nu } $.  If  the cancellation of $d =4$ anomalies in  compactified Horava-Witten theory  mirrors what happens in the weak case, ( as for example  the 
cancellation of $ d=10 $ anomalies does), then  $c$ should be  proportional to 
the pure $U(1)^3 $  anomaly and  mixed $U(1) $ -gauge and  $U(1) $ -gravitational 
anomalies of the compactified theory. In fact the coefficients of all 
three anomalies must be proportional to each other (if they are non
zero) , 
if  one is to achieve complete cancellation by an appropriate $U(1) $ 
gauge transformation of  the field $S$, 
as implied by the invariance of  the K\"ahler potential $L_K $.  Such 'universality '
conditions have been described in the context of orbifold compactifications of 
the weakly coupled heterotic string in \cite{kob}, and they provide quite a
strong  constraint on the kind of embeddings that can give rise to anomalous
$U(1) $'s. Assuming an unbroken gauge group 
of the form $U(1)_A \, \times \Pi_a G_a $ with $G_a $  semi-simple, 
the form of this universality  relation is 
\beq\label{eq:univ}
\frac{1}{k_a} \, Tr_{G_a} ( T( R )  Q_A ) \, = \, \frac{1}{3} Tr Q^3_A \, = \, 
\frac{1}{24} Tr Q_A 
\eeq
where $2 T( R) $ is the index of the representation $R$  and $k_a $ 
its level. Also, the  $U(1)_A  $
generator  $Q_A $ has been rescaled  such that the level $k_A \, = \, 1 $.

In studying   non-standard embeddings and the possibility of anomalous
$U(1)$ symmetries of the $d=4 $ theory, there are at least two basic 
situations to consider. 
The first  (case I) is when  the abelian parts of the background field 
strengths
$F^{(i)}_{a \bar{a}} $ on $X$ are in a direction orthogonal to the 
$U(1)$'s under consideration, and the second  (case II) is when the background values are taken to lie
in the same directions as the $U(1)$'s.  
On top of this, in the second of these cases  we have the possibility of 
mixing between $U(1)$'s (assuming there is more than one 
$U(1) $ factor ),  depending on the type of embedding
chosen. 
The  emebdding defined in (\ref{e:32}) is an example of the case II type discussed above, with mixing. 

To have a better idea about anomaly cancellation in $M$-theory in four dimensions,
before discussing potential $U(1)$ anomalies in case II, 
we shall first 
study case I, which is conceptually easier. 
Although we do not expect to find anomalous $U(1)$'s in this case since the 
embedding does not mix between hidden and observable sectors and we would not
 know how to cancel such an anomaly, it  is instructive to show this 
formally, and moreover it turns out that the resulting formulae can easily 
be extended to case II type embeddings. 
Furthermore, in both cases I and II, for consistency, we should find 
that  $d=4$
triangle diagrams vanish precisely when the corresponding Green-Schwarz terms do, 
and if they are non-vanishing (which is the most intersting case),  so must the Green -Schwarz terms. This is what we shall 
verify in the remainder of this section.

 With this in mind,  we start, as promised, by considering 
case I type embeddings where
the $E_8  $ and $E_8'$ gauge symmetries arising on each wall  break via the
subgroups
 $G \times U(1) \times J $ and $ G' \times U(1)' \times J' $ respectively.
The background gauge fields $F_0^{(i)} $ on $X$ are assumed to take values in the subgroups $J $ and $J' $ for $i=1,2$, which are not necessarily semi simple. 
This background will give rise to an unbroken gauge group $G \times U(1)\times G'\times U(1)' $, although whether the $U(1)$ factors are further broken depends on the anomaly analysis. 
We  write the decomposition of the adjoint representations $248 $ and $248'$ of respective $E_8 $ 's as 

\beq \label{e:39}   
248 \quad  = \quad   \sum_{a_1} \, L_{a_1} \otimes y_{a_1} \otimes Q_{a_1} \, ,  \qquad    
248' \quad  = \quad   \sum_{a_2}\,  L_{a_2} \otimes y_{a_2} \otimes Q_{a_2}
\eeq
where  $a_1 , a_2 $ runs over the number of 
irreducible representations in the respective decompositions, 
with $L_{a_i} \, ; Q_{a_i} $ defining irreducible representations 
of $G, G' \, ; J, J' $ for $i =1,2$, 
while $y_{a_i}$ define the corresponding $U(1), U(1)' $ charges.

Defining $Y, Y' $ to be the generators of $U(1) , U(1)' $ and
following the reduction of the $CGG$ term  in $d = 11 $ outlined 
above, one finds  the terms 
$c_1  \, \int d^4x \, B \wedge F_1 $ and   $c_2  \, \int d^4x \, B \wedge F_2 $  in four dimensions, where $F_1 , F_2 $ are the respective two form field strengths of
$U(1) $ and $U(1)' $. The coefficients $c_1,\, c_2 $ are given by 
\beqa \label{e:40}
c_1 \quad & = & \quad   \,  \frac{3 \rho_0 }{2 \pi \kappa^2 } (
\frac{-\sqrt{2}}{3456} )
 (\frac{\kappa}{4 \pi } )^{4/3} \int_X \frac{1}{30} Tr( Y F_0^{(1)} )
 \wedge \left (
\frac{1}{30} Tr( F_0^{(1)}
 \wedge F_0^{(1)}) \right .
 \cr
&&\cr 
&-&\frac{1}{60} \left . Tr ( F_0^{(2)} \wedge F_0^{(2)}) - \frac{1}{4} tr ( R \wedge R) \right )  \cr
&&\cr
c_2 \quad & = & \quad   \,  \frac{3 \rho_0 }{2 \pi  \kappa^2} (
\frac{-\sqrt{2}}{3456} )
 (\frac{\kappa}{4 \pi } )^{4/3} \int_X \frac{1}{30} Tr( Y' F_0^{(2)} ) 
\wedge \left (\frac{1}{30} Tr ( F_0^{(2)} \wedge F_0^{(2)}) \right .
 \cr
&&\cr 
&-&\frac{1}{60} \left .
 Tr ( F_0^{(1)} \wedge F_0^{(1)}) - \frac{1}{4} tr ( R \wedge R) \right )  
\eeqa

where in (\ref{e:40})  $Tr $ means trace in the adjoint  
representation $248 $ or $248'$ , and we recall that it is the
coordinate 
radius $\rho_0 $ that  appears in these expressions.  Both these 
coefficients vanish if we specialize to the standard embedding.  
The second one vanishes since in this case $F_0^{(2)} \, = \, 0 $ , 
whilst the first one vanishes because  the 
$[ Y, F_0^{(1)}] \, = \, 0 $  allows the decomposition  
$Tr(YF^{(1)} )  \, = \, 
\sum_{a_1} dim(L_{a_1}) \, tr_{ y_{a_1}} (Y)  tr_{Q_{a_1}}( F_0^{(1)}
)  $, which 
is vanishing since $ tr_{Q_{a_1}} ( F_0^{(1)} )  \, = \, 0 $ for the standard embedding. Whether in the case I type embeddings we are considering, 
$c_1 , c_2 $ are always necessarily vanishing is not totally clear. Certainly, 
if we choose embeddings in such a way as the generator $Y$ is orthogonal 
to  $F^{(1)} $, i.e  $Tr( Y F^{(1)} ) \, = \, 0 $, and  in a similar manner $Y' $ orthogonal to $F^{(2)} $, then indeed these coefficients are vanishing. 
This was the case in the particular example, and for the particular choice of 
nonstandard embedding  considered in \cite{dist}, in  (weakly coupled) 
$E_8 \times E_8 $ heterotic (2,0) compactifications.

For consistency, it should be the case that  $c_1,  c_2 $ must also be
proportional to the  coefficients of the various $U(1) $ anomalous 
triangle diagrams
in $d=4 $. Checking this will provide a good test of the explicit form
of the topological integrals over the C-Y space $X$   in terms of which 
$c_1, c_2 $ are defined.
Consider, for example, the mixed  $U(1)-GG $ or 
$U(1)'-G'G' $  anomalies.  The corresponding anomaly coefficients  
$ I_{UGG}$ and  
$I_{U'G'G'}  $  may be written as 
\beq\label{e:41}
I_{UGG} \, = \,  \sum_{a_1}  n_{L_{a_1}, y_{a_1}} \, tr_{ L_{a_1}, y_{a_1}} ( YTT) \qquad , \qquad  
I_{U'G'G'} \, = \, \sum_{a_2}  n_{L_{a_2}, y_{a_2}} \, tr_{ L_{a_2}, y_{a_2}} ( Y'T'T')
\eeq
where   $ n_{L_{a_i}, y_{a_i}} $ represents the number of chiral 
fermions transforming in the  corresponding irreducible
representations  
labelled by  $ (L_{a_i}, y_{a_i} )$, 
and $ T , T' $ denote  the generators of $G , G' $.  These can be expressed , through the use of Index theorems,  in terms of various topological integrals over $ X $ (e.g.
see \cite{gsw} suitably generalized to the present case ):
\beq\label{e:42}
n_{L_{a_i}, y_{a_i}} \quad = \quad  \frac{1}{48 (2 \pi)^3 } \int_X  \, tr_{Q_{a_i}} \left (  F_0^{(i)} \wedge F_0^{(i)} \wedge F_0^{(i)} \right )   - \frac{1}{8} 
tr_{Q_{a_i}} (F_0^{(i)}) \wedge tr(R \wedge R) 
\eeq

If  $W , W' $  represents a generator of  $G \times U(1)  \, , \, G' \times U(1)' $ 
and  $ Z,  Z' $ that of $J \, , \, J' $, then it follows that  $ \sum_{a_1} tr_{L_{a_1}, y_{a_1}} ( W ) tr_{Q_{a_1}} ( Z )\, = \, Tr (WZ) $ with a similar relation for $W'$ and $Z' $.  Using this fact, one can obtain 
\beqa\label{e:43}
I_{UGG}  & = &  \frac{1}{48 (2 \pi)^3 } \int_X \left ( Tr(YT^2 (F_0^{(1)})^3 )  - 
\frac{1}{8} Tr (YT^2 F_0^{(1)} ) \wedge tr (R^2)  \right ) \cr
&&\cr 
&&\cr
I_{U'G'G'}  & = &  \frac{1}{48 (2 \pi)^3 } \int_X \left ( Tr(Y'T'^2 (F_0^{(2)})^3 )  - 
\frac{1}{8} Tr (Y'T'^2 F_0^{(2)} ) \wedge tr (R^2)  \right ) 
\eeqa
where in these equations we use the notation that $Tr(A^m ) \, = \, Tr ( A
\wedge A \wedge A ....\wedge A) $. To proceed we make use of the well known identities 
$TrF^6 \, = \, \frac{1}{48} TrF^2 \, TrF^4 - \frac{1}{14400} \,  (TrF^2)^3 $ and $ TrF^4 \, = \, \frac{1}{100} \,  (TrF^2)^2 $. Although these identities involve the trace $Tr$ in the adjoint of $E_8 $
it is an important  fact (which we shall return to later) that the first of these identities also holds for traces over the adjoint of $E_8 \times E_8 $.  Using these identities 
with the choice $F \, = \alpha Y + \beta T + \gamma F_0^{(1)} $ or $ 
F = \alpha Y' + \beta T' + \gamma F^{(2)} $ and expanding the various monomials
we obtain 
\beqa\label{e:44}
I_{UGG}  & = &  \frac{1}{48 (2 \pi)^3 } \frac{1}{60 \cdot 1200}\int_X \left
  ( Tr(T^2)
 Tr(YF_0^{(1)})\wedge  ( Tr(F_0^{(1)2} )  -  Tr (F_0^{(2)2} ) ) \right ) \cr
&&\cr 
&&\cr
I_{U'G'G'}  & = &  \frac{1}{48 (2 \pi)^3 } \frac{1}{60 \cdot 1200}\int_X \left ( Tr(T'^2 ) Tr(Y'F_0^{(2)} ) \wedge ( Tr (F_0^{(2)2}) )  -  Tr (F_0^{(1)2} )  )   \right  ) 
\eeqa
In obtaining these equations we have also made use of the semi-simple properties of 
the generators  $T , T' $, as well as the Bianchi identity $dH \, = \,
 1/30 (Tr F_0^{(1)\,2}   + Tr F_0^{(2)\, 2} ) - tr R^2 $.  Note that  
the term $dH$ does not contribute since an integration by parts gives
zero.
This follows from the fact that 
$Tr(YF_0^{(1)}) $, when decomposed into traces over the irreducible
representations  $Q_{a_1} $, gives 
an effective U(1)  valued 2-form, which  satisfies an abelian  Bianchi
 identity (the same holds for  $Tr (Y' F_0^{(2)} $). 

We can now compare the expressions for $I_{UGG}$ and  $I_{U'G'G'} $ 
and those for  $c_1$ and  $c_2 $ in (\ref{e:40}). First we notice that  we can 
remove the  $\rho_0 $ dependence in the latter equations  by using  the 
equation (\ref{e:corr}) relating M-theory and string parameters.
Furthermore, again using the Bianchi identity involving $dH$ and the 
integration by parts argument mentioned above,  one can indeed verify that the 
forms of 
the expressions for $c_1, c_2 $ are in agreement with those of 
$I_{UGG} , I_{U'G'G'} $.

Having established this connection for case I type embeddings, we now consider
case II type.  After some thought, it is apparent that the corresponding 
coefficients $c_1 , c_2 $ as well as their relation to the $U(1) - G G $ and $U(1)' -G'G' $ anomalies will be given by the same expressions, but now $Y, Y' $
are identified with the $U(1) $ generators of $J$ and $ J' $. 
Then, in contrast to  case I embeddings, it is clear that $c_1 , c_2 $ will not be vanishing in general, since $Tr (Y F^{(1)} ) , Tr ( Y' F^{(2)} )  $ 
will be 
non-zero, and we have the possibility of  a surviving anomalous $U(1)$.

At this stage it remains only a possibility, because whilst  it is a 
necessary condition that, for an anomalous $U(1) $ symmetry to exist, 
the coefficients of the corresponding 
$B \wedge F$ terms must
be related to the various anomaly coefficients as we have seen above, 
it is not sufficient. This is because, as we discussed earlier, to 
some of these potentially anomalous
$U(1)$'s the Higgs mechanism can be  applied which involves  
the $H$ field kinetic energy term improved by Chern-Simons terms, 
and this can happen precisiely in the case II embeddings we are discussing. 
This is the case for the $U(1)$s which belong to the structure group of 
the gauge vacuum bundle. These $U(1)$s can be anomalous without being 
coupled to both walls. The reason is precisely that in this case the Higgs
mechanism involves 
model-dependent axions through the improved kinetic term for the 
antisymmetric tensor field. In this paper, motivated by 
phenomenological applications, 
we are mainly interested in an  anomalous $U(1)$ for which the Higgs mechanism
involves  
the model-independent axion. 
As we have stressed earlier, (and which motivated the choice of mixed
embedding in (\ref{e:32}) ) such an anomalous $U(1)$ must 
couple to both walls. If it does not then it should not be anomalous.
Indeed it is a non-trivial check on the ideas presented in this section, that 
for example, if we imagine an embedding similar to the one in (\ref{e:32}), but now
defined purely in terms of line bundles on a single wall (say we are 
switching on the background which mixes two $U(1)$s living on the 
first wall), with field strengths 
$F_{(1) a \bar{a} } \, , \, F_{(1) a \bar{a}}' $, then the surviving $U(1)$
should not be anomalous.  This is an example of a  case II type embedding, 
only we have mixing 
between  two $U(1) $'s , both of which lies in the same $E_8 $ factor. 
Whilst the formula presented earlier have not formally covered 
this possibility, 
it is easy to extend them to do so, simply  reinterpreting formulae 
(\ref{e:40}). 
For the resulting 
coefficients  $c_1$ and  $ c_1' $ of the $B \wedge F$ terms one obtains
\beqa\label{e:80}
c_1 \quad & = & \quad   \,  \frac{3 \rho_0 }{2 \pi \kappa^2 } (
\frac{-\sqrt{2}}{3456} )
 (\frac{\kappa}{4 \pi } )^{4/3} \int_X \frac{1}{30} Tr( Y F_0^{(1)} ) \wedge \left (\frac{1}{30} Tr( F_0^{(1)}
 \wedge F_0^{(1)}) \right .
 \cr
&&\cr 
&-&\frac{1}{60} \left . Tr ( F_0^{(2)} \wedge F_0^{(2)}) - \frac{1}{4} tr ( R \wedge R) \right )  \cr
&&\cr
c_1' \quad & = & \quad   \,  \frac{3 \rho_0 }{2 \pi  \kappa^2}
 ( \frac{-\sqrt{2}}{3456} ) (\frac{\kappa}{4 \pi } )^{4/3} \int_X  
\frac{1}{30} Tr( Y' {F'}_{0}^{(1)} ) 
\wedge \left ( \frac{1}{30}Tr ( F_{0}^{(1)} \wedge F_{0}^{(1)}) \right .\cr
&&\cr 
&-&\frac{1}{60} \left .
 Tr ( F_{0}^{(2)} \wedge F_{0}^{(2)}) - \frac{1}{4} tr ( R \wedge R) \right )  
\eeqa

For the embedding defined by 
\beq \label{e:70}
 \left ( \cos (\theta) F_{(1) a \bar{a}} + \sin (\theta) 
F_{(1) a \bar{a}}' \right ) 
=  {\cal F}_{a \bar{a}}'
\eeq
it  is easy to see that the corresponding coefficients of  the  surviving 
 $B \wedge F$   indeed vanish.
One may verify 
from  (\ref{e:80})  that they are  proportional to the 
combination $-sin (\theta ) F_{(1) a \bar{a}} 
+ cos (\theta ) F_{(1) a \bar{a}}' $, which by assumption is vanishing . 
Of course, the coefficient of the 
structure $U(1)$, whose gauge boson is eaten up by the Higgs mechanism, is 
nonzero. However, as we remarked before this is 
not the $U(1)$ of the type we are interested in. It decouples  
from the massless spectrum, hence its anomaly is harmless.  

By the same reasoning, we should find that if we  choose an embedding like
(\ref{e:32}) which mixes  fields on different walls, then the 
coefficient should be non-vanishing in general.
Following the same logic as before, we  find that
in $d= 4 $ the term $ c' \, \int B \wedge F  $ arises  from the $CGG $ term reduction, 
where $F$ is the orthogonal combination  of  $F_1 , F_2  $  that defines the unbroken 
$U(1)$, and $c' \, = \, cos (\theta ) c_2 - sin (\theta ) c_1 $ . Next, 
we have to use  equations  (\ref{e:40}) ( we stress that in 
these equations 
$Y, Y' $ are taken to define the $U(1) $ generators in  $J, J' $ ),  appropriate for each wall. An important point is that using the $dH$ Bianchi identity, 
we see  that the 4-forms multiplying the $Tr(YF_1^{(1)} ) $
and  $Tr(Y'F_2^{(2)} ) $ terms in (\ref{e:40}) are equal but with opposite 
sign. From this we see that 
the combination  $cos (\theta ) F_{2}^{(2)} + sin (\theta ) F_{1}^{(1)} $ occurs in 
the coefficient $c'$, which is equal to $2 \, \sin(\theta) \, F_{1}^{(1)}$
upon using the constraint discussed below (\ref{e:70}. This 
is 
non-vanishing as implied by definition of the embedding (\ref{e:32}). 
The crucial sign flip seen in $c'$, whose origin lies in the 
Bianchi identity, is responsible for providing (in general) 
an anomalous $U(1) $
in $d=4 $. This argument also illustrates the idea that such an anomalous $U(1)$ must couple to both walls. 
If we put $\theta=0$ (or $\pi/2$), which confines embedding to a single wall, 
then $c'$ vanishes. 

We now make some comments concerning the mass scale entering the 
Green-Schwarz term of the anomalous $U(1)$ we obtained above. 
To do this it is convenient to  solve the $B$-field equations of motion in the frame defined
earlier  
in (\ref{e:fr}) (and recalling  $g_{55}\, = \, e^{2 \, \gamma } $ ), which allows one to make contact with the axionic pseudo-scalar field $b$ 
\beq\label{e:100}
 {\star H}_{\mu} \, = \, \frac{e^{\gamma- 6 \beta}}{\sqrt{ g^{(6)} } }  \partial_\mu b    
\eeq
where ${\star H} $ is the dual of the 3-form field strength of $B$.
Substituting this solution back into the effective four dimensional action,
one can verify that  the  Green-Schwarz term expressed in terms of the variable
$b$ is consistent with the expansion of the K\"ahler potential
$K \, = \, - M^2_{Pl} \, ln ( S + \bar{S} - c V ) $. Thus,  we find the    Green-Schwarz term  $\sim \,  g^2 \, M^2_{Pl} \, \partial_\mu b  A^{\mu } $
where $A^{\mu} $ is the gauge field of the anomalous $U(1)$ . 
Here $g^2  \, = \, g^2_1 \, = \, g^2_2 $ is   
 the lowest order term in the $\kappa $ expansion of $g^2_{mixed} $,  where we note that  it is only the threshold corrections  (which are proportional to $\kappa^{-4/9} $)
which distinguish $g_1 $ from $g_2 $ in the definition of $g^2_{mixed} $.  (Recall
that   the factor of $M^2_{Pl} $ already contains a factor of $\kappa^{-4/9} $ ) 

Before ending this section,  we make some comments  concerning  
the connection between the above results and those of the weakly coupled case. 
We expect that in this case the generation of $ B \wedge F $ terms in four dimensions
comes from the reduction of the $d=10$ Green-Schwarz terms
 \cite{green}. The  coefficient of  such a term should once again be  proportional to the various $U(1)$ triangle anomalies  in four dimensions. We can derive expressions for the latter in terms of 
the index of the Dirac operator on $X$ where now  we  are considering the full $E_8 \times E_8 $ gauge group (as opposed to separate $E_8 $ factors on each wall in the
M-theory case ). However it turns out that the reduction of the  GS terms gives 
expressions for the coefficient of $B\wedge F $ that  mirror the expressions 
of $c_1 , c_2 $  calculated above, in the strongly coupled case.  Therefore the 
$d = 4 $ anomaly analysis must be mirrored likewise.  
This follows from 
 the nontrivial property of the $TrF^6 $ identity which we mentioned earlier, 
namely that it is satisfied for  $E_8 $ and $E_8 \times E_8  $, separately. 
This  is an important key to understanding why the anomaly analysis, in both 
$d=10 $ and as we have seen in this paper,  $d = 4 $,  are mirrored in the weak and strong coupling cases.
      
\section{Gauge Couplings in 
Nonstandard Embeddings}

In this Section we shall discuss the issue 
which has direct phenomenological relevance: the correlation 
between the choice of embedding, hence of the structure of the 
unbroken low energy symmetry group, and the values 
of the unified gauge couplings on the fixed hyperplanes. 
Let us recall the result 
for the gauge coupling difference,
 originally  
given by Witten \cite{strw} and also following from Section 3: 
\beqa \label{e:45}
\alpha_{h}^{-1} - \alpha_{v}^{-1} &=& \frac{2}{ (4 \pi \kappa^2)^{2/3}}
 ( V_h - V_v ) = \frac{ \pi \rho_p\, s_i }{(4 \pi)^{1/3} \kappa^{2/3}}
\frac{1}{8 \pi^2} \int_X \omega \, \wedge \, (\, trF^{(i)} \wedge F^{(i)}
 - \frac{1}{2} \, trR \wedge R \,) \nonumber \\
&& 
\eeqa
where $s_i \, = +1, -1 $ for $i = 1, 2 $ respectively.
We stress again, 
that this result is independent of the particular 
embedding.
The split of the couplings given above
is of the relative order $o(\kappa^{4/3})$ 
with respect to the gravitational part of the Lagrangian, and it has 
the interpretation  of the difference between  volumes 
of the two C-Y spaces  on the hidden and visible walls (with the 
proportionality coefficient $2 \pi (4 \pi \kappa^2)^{2/3}$.

Let us 
rewrite (\ref{e:45}) in terms of string units, i.e. in terms of $\alpha'$
\beqa \label{e:46}
\alpha_{h}^{-1} - \alpha_{v}^{-1} &=& \frac{s_i}{8 \pi^2 \alpha' }
\frac{1}{8 \pi^2} T  \int_X \omega \, \wedge \, (\, trF^{(i)} \wedge F^{(i)}
 - \frac{1}{2} \, trR \wedge R \,) 
\eeqa
To better understand this expression  one should write down the integrand in the explicit manner
\beq 
\omega \wedge tr(F \wedge F) = g^{a \bar{a}} g^{b \bar{b}} tr ( F_{a \bar{a}}
F_{b \bar{b}} ) - g^{a \bar{a}} g^{b \bar{b}} tr ( F_{a \bar{b}}
F_{\bar{a} b} )
\eeq
Using the Einstein-K\"ahler equations given in the preceding section 
one easily finds  that
\beq
\int_X \omega \, \wedge \,  tr(F^{(i)} \wedge F^{(i)}) = 
-\frac{1}{2} \, \int_X tr F^{(i)}_{ij} F^{(i)\,ij}
 = - 8 \pi^2 \, n_{F\, i}\, \leq \, 0
\eeq
The same steps can be repeated for the gravitational part of the integrand
\beq
\int_X \omega \, \wedge \,  tr(R \wedge R) = 
-\frac{1}{2} \, \int_X tr(R_{ij} R^{ij}) = - 8 \pi^2 \, n_{R}\, \leq \, 0
\eeq
Hence the integral that gives the difference between the unified couplings 
is 
\beq 
\int_X \omega \, \wedge \,  (\, tr(F^{(i)} \wedge F^{(i)})
- \frac{1}{2} \, tr(R \wedge R)\,) = - 8 \pi^2 \, ( \, n_{F\, i} - \frac{1}{2} 
\, n_{R}\, )
\eeq
The integrability conditions for the equations of motion give constraint on 
the instanton numbers\footnote{The configurations we use here 
fulfill the Yang-Mills equations of motion and Einstein equations which justifies the term instanton.} \cite{gsw}
\beq 
 n_{F\, 1} +  n_{F\, 2} = n_{R} 
\eeq
It is important to note that all the above instanton numbers are 
positive (some 
of the gauge ones may be zero). In fact,    
from the observation that on any Calabi-Yau manifold the standard embedding 
must be possible, it follows that Calabi-Yau spaces always have 
$n_R \, > \, 0$. Moreover, in the case of a general embedding $n_{F\, i}$ 
must be positive (or zero) for any bundle in the direct sum, 
since the K\"ahler-Einstein equations must be fulfilled for each 
$V_i$ separately. 
In terms of the instanton numbers and in units in which $2 \alpha' =1$
the difference of inverse couplings is  
\beqa \label{e:47}
\alpha_{h}^{-1} - \alpha_{v}^{-1} &=& - \frac{s_i}{4 \pi^2 }
 T_r (n_{F \, i} - \frac{1}{2} n_R ) 
\eeqa
Let us take standard embedding first. There $n_{F \, v} = n_R$, and
\beqa \label{e:48}
\alpha_{v}^{-1} - \alpha_{h}^{-1} &=& \frac{1}{8 \pi^2 }
 T_r n_R  
\eeqa
This particular embedding gives the specific gauge group structure 
$E_{6 (v)} \times E_{8 (h)}$. We stress that since $n_R$ has  
positive sign, there is no way of changing the sign of $\alpha_{v}^{-1} - 
\alpha_{h}^{-1}$. 
For individual gauge couplings we have 
\beq \label{vt}
\alpha_{v}^{-1} = 4 \pi \,(S_r + \epsilon \, T_r), \; \alpha_{h}^{-1} = 
4 \pi \, (S_r - \epsilon \, T_r );\;
\epsilon = \frac{n_{F \, v} - \frac{1}{2} n_R}{32 \pi^3}
\eeq
Let us look for another model, which has $\epsilon ' = - \epsilon$. 
This means that $n_{F \, v} - \frac{1}{2} n_R = - 
n_{F \, v} '  + \frac{1}{2} n_R '$. 
Of course, one of the solutions to this equation is the `complementary'
sector from the original standard embedding. This is the only solution if 
one demands that the CY metric does not change, i.e. if $n_R = n_R '$. 
If one allows for  other metrics, then one can have more 
interesting solutions. Let us also note  that it is possible to obtain 
the situation where  $\epsilon = 0$, see \cite{step} for $N=2$ examples,
 however we shall not pursue in the 
present paper any specific model in detail.

As far as general layout is concerned, there are two broad classes
of models: the ones where the visible sector belongs to the more weakly 
coupled , say weakest, part of the unbroken gauge group, 
and the ones where the visible sector is the most strongly coupled one. 
We shall always assume that none of the nonabelian groups from the 
visible sector is of the mixed type. These two distinct unification routes 
which open up in the general embedding case shall be discussed below.

\subsection{Weaker Obervable Sector} 

We assume here that we live on the 
hyperplane where the C-Y volume is larger, hence the unified coupling is
smaller than the one on the other hyperplane. The obvious constraint is 
that the hidden coupling is smaller than infinity, see 
(\ref{vt}). This requirement, formulated in the 11d supergravity frame
used originally by 
Witten,
leads to the notion of the critical radius. This is the physical 
distance between the planes that corresponds to vanishing volume of the 
C-Y space localized on the hidden plane. 
Using linear approximation for the variation of the volume along $x^{11}$,
one can  obtain the value of the physical critical radius 
$(\pi \rho_{c})^{-1} = 0.8 \times 10^{15} \, GeV$. 

One should note that in the 5d supergravity frame 
the presence of the  critical radius manifests itself 
in a different way than in the 11d canonical frame, since in the equation for 
$M_{Pl}$ there is no modulus $S$.
In the 5d canonical frame  we see the critical radius 
once we allow the $G_N$ to vary  while $g_{GUT}$
and $M_{GUT}$ are fixed at their observed values, and one takes a specific,
fixed by an embedding, value of $\epsilon$. Then one can  find  that 
in order not to leave the field theoretical domain, the physical 
radius of the eleventh dimension must be smaller than $\rho'_{cr} =
\rho_0 /(2 \,  g^{2}_{GUT} \, \epsilon )$.

Another way to explore the constraint  of  finite $g^{2}_h$
is to allow the unified coupling constant to vary
while keeping the other parameters fixed at realistic values. 
One obtains the condition
\beq \label{ulm}
\alpha_v \, < \, 
4 \pi^{3/2} \frac{1}{\epsilon^{3/4}}
 (M_{GUT} /M_{Pl})^{3/2}
\eeq
This  puts some restrictions on model building. To be more 
specific, let us take the example of K3 fibration C-Y manifolds 
discussed in \cite{klemm}. Guided by this example we see that a typical value 
of $\epsilon$ is $\epsilon=O(10)/(32 \pi^3)$. This leads to the condition 
$\alpha_{v} \, < \, 0.047$. The upper limit for the standard embedding value 
$\epsilon_s = 3/(8 \pi^3)$ is $\alpha_{(s)\,v} \, < \, 0.041$. 
This shows that while the observable unification coupling 
$\alpha_{GUT} = 0.04$
fits marginally within the limits, accommodating
the scenarios of `strong' unification \cite{ross}, with 
$\alpha_{GUT} \approx 1$, within this branch of 
$M$-theory models is not possible.
The problem is underlined by the fact that the hidden sector coupling
is, by assumption, even stronger. Actually it grows with the growing 
visible coupling according to  
\beq
\alpha_{h} =
\left ( \frac{1}{ \alpha_v} - \epsilon ( \alpha_v )^{1/3} 
(\frac{M_{Pl}}{M_{GUT}})^2
\frac{1}{ 2^{2/3}4 \pi^2} \right )^{-1}
\eeq
The value of $ \alpha_{h}$ becomes larger than one 
at $\alpha_v =0.045$ for 
a generic embedding and slightly above $\alpha_v=0.04$ for the 
standard embedding. 
This means in particular  that, generically, in this type of scenarios 
there is 
not much space for running of the gauge coupling in the hidden sector. 
In consequence, it is difficult to develop a condensation scale which would 
be hierarchically smaller than the unification scale, as needed for 
realistic supersymmetry breaking. The condensation scale is given by the 
formula:
\begin{equation}
\Lambda_c=M^{\prime}_{GUT}e^{- 1\over {8\pi b_{(h)0}\alpha_h}}
\label{e:mgut}
\end{equation}
where $M^{\prime}_{GUT}=M_{GUT}(\alpha_h/\alpha_v)^{1/6}$ is the unification
scale in the hidden sector. 
For the standard embedding with the pure
$E_8$ in the hidden sector ($\alpha_h=0.97, \;\;b_0 = 0.57$) one obtains
$\Lambda_c = 1.59 \, M_{GUT}$.  
Changing the unifying gauge group so that 
$0.1 \, < \, b_0 \,< \, 0.57$ and for 
embeddings with generic value of $\epsilon$ we get 
$0.18\, M_{GUT} \, 
< \Lambda_c \, < \, 0.92 \,  M_{GUT}$.  
Although nonstandard embeddings lower the value of the condensation scale 
by about an order of magnitude, 
this is still clearly above the border line of giving the proper gaugino 
condensation scale. Further decrease of the 
condensation scale would require additional matter fields, which would 
decrease the coefficient $b_{(h)0}$.

\subsection{Stronger Observable Sector}

A very interesting class of models arises, when one identifies the 
visible sector with the part of the unbroken gauge group which is 
most strongly coupled (this possibility has been independently considered
in \cite{benakli}).  
To obtain 
relations involving $\epsilon$, $S$ and $T$ we solve 
again the strongly coupled 
string relations (\ref{bas})
from the Section 2.1, substituting there this time 
$1/\alpha_{GUT} = 4\pi(S_r -  \epsilon \, T_r )$.

Obviously, there is no critical radius and 
no critical value of $\epsilon$ in the sense that the C-Y volume on the hidden
hyperplane is always larger than that on the visible hyperplane
(we remind that to orient our discussion we are assuming 
$\epsilon > 0$). Hence 
it 
never becomes smaller than $(\alpha')^3$. 

However, there are limits on $\alpha_{v}$ or/and $\epsilon$ 
coming from two places. Firstly, the string coupling 
$g_{s}^{2}$ should remain larger than unity, to stay within the realm of 
the Horava-Witten model, and away from the transition limit between 
weakly and strongly coupled string. Thus, we demand  
$S/T^3 \, < \, 1$\footnote{This condition is always fulfilled if the observable sector is the weakest one.}. 
If we treat $\alpha_{v}$ as a parameter, then this translates 
into the condition
\beq
\frac{1}{\alpha_v} + \epsilon \alpha_{v}^{1/3} \left ( \frac{M_{Pl}}{M_{GUT}} \right )^{2} 
\frac{1}{ 2^{11/3} \pi^{2}} \, < \, 
\alpha_v \left ( \frac{M_{Pl}}{M_{GUT}} \right )^{6} 
\frac{1}{ 2^{15} \pi^{8}}
\eeq
which is fullfilled with a wide margin for typical values of $\epsilon \approx
O(10)/(32 \pi^3)$. 

The other, more interesting limit comes from the requirement that it is
hierarchical 
dynamical 
supersymmetry breaking on the hidden wall which is responsible 
for soft masses 
in the visible sector. In general, for this to work, one 
needs a dynamically 
generated condensation 
scale in the hidden sector to be $10^5 \, GeV\;< \, \Lambda_{cond}
\, < \; 10^{13} \, GeV$ (a mass range which interpolates 
between gauge mediated and gravity mediated supersymmetry breaking models). 
Using the analog of eqs.(69) and (\ref{e:mgut}) it is straightforward to work 
out the relation between the hidden condensation scale and $\epsilon$.
Taking  a typical value of $\epsilon=0.01$, one obtains 
$1.4 \, 10^{-8} \, M_{GUT} \,<\, \Lambda_{c} \, < \, 2.9 \, 10^{-7} 
\, M_{GUT}$
for $0.04 \, < \, \alpha_{v} \, < \, 0.11$, and then the value of the condensation scale becames smaller again achieving $2.4 \, 10^{-11}\, M_{GUT}$ at 
$\alpha_v =1$.
In the above estimate we have put the value of one-loop beta-function 
coefficient to $0.1$. In principle, it is possible to increase the 
condensation scale by increasing $b_{(h)0}$. For example, 
$b_{(h)0}=0.25$ would give $\Lambda_{c}= 7 \, 10^{-3} \, M_{GUT}$
at $\alpha_{v}=0.04$.

It is interesting to find out how large the hidden wall Calabi-Yau 
might be. It turns out it cannot be too large, or in other words the mass 
of the heavy Kaluza-Klein modes is as large as in the visible sector. 
The mass of the hidden Kaluza-Klein modes in terms of $\epsilon$ is given as:
$M_{h\, KK} = \frac{M_{GUT}}{(4 \pi \alpha_{GUT})^{1/6}} (S_r + 
\epsilon T_r )^{-1/6}$.
For  the typical choice  of  $\epsilon$ motivated by the $K3$ fibrations C-Y 
examples,
the smallest value one can get 
is $M_{h\, KK} = 0.51 \, M_{GUT}$ corresponding to 
$\alpha_{GUT} = 1$\footnote{And 
$M_{h\, KK} = 0.91 \, M_{GUT}$ for $\alpha_{GUT} = 0.04$.}. 

In the presently discussed scenario where the coupling of 
the observable group is 
the strongest, one can raise it up towards the nonperturbative region,
without violating any bound, this way providing a realization of the ideas
of strong unification advocated in \cite{models}.

It is also worth of pointing out, that 
the race-track gaugino condensation scenarios 
also need nonstandard hidden sector with hidden matter in order to work 
properly \cite{b9}.

Finally, we observe that non-standard embedding $M$-theoretical models 
might  naturally be  good places for the realization of the scenario 
of supersymmetry breaking and moduli stabilization proposed in \cite{b1}.

\section{Summary}

In this paper we have discussed theoretical and phenomenological aspects of
general embeddings in M- theory compactified on $S^1/Z_2$. Going beyond the
standard embedding discussed so far in the literature is interesting for 
several reasons. We focused our attention on the existence of  anomalous
$U(1)_A$ and on the behaviour of the 
gauge couplings in the non-standard embedding
scenarios.

As a necessary technical component of this discussion we first generalized
the standard embedding result for the threshold corrections  to the gauge
kinetic couplings  to non-standard embeddings.
For  general embeddings,   
we have formulated the effective four dimensional $N=1$ supersymmetric 
theory by solving the Bianchi identity and equations of motion for the 
antisymmetric tensor background along the eleventh (fifth) dimension,
and integrating out the explicit $x^{11}$ dependence.   
The  solution for that background, and implicitly also 
for the metric, is given in terms of 
massless and {\em massive } eigenmodes of the Laplacian on Calabi-Yau 
space. Using this result, we have shown that,  
when one considers only the effective 
couplings between the {\em massless} fields, the form of threshold 
corrections 
is  the same as for the standard embedding in the sense that in the expression 
$ 1/ g^{2}_{(1),(2)} = Re(S + (\pm \epsilon_{i\, (1), (2)} ) T_i )$ 
(where $i=1,\ldots,h{(1,1)}$) the individual coefficients 
$\epsilon_{i\, (1),(2)}$ are related through $\epsilon_{i\, (1)}=
- \epsilon_{i\, (2)}$. An easy way to obtain this result is to consider
the axial part of the corrections which follow from the 
reduction of the $C \wedge G \wedge G$ 
term. 

A particular class of embeddings which we have discussed in detail is the one
in which gauge interactions  `mix' the two walls, that is 
under which fermions on {\em both} hyperplanes are charged. 
We explicitly give such a construction, and point out that this is the 
way the anomalous $U(1)_A$ gauge group can arise 
in $M$-theory models. We  show how the four dimensional Green-Schwarz term
$B\wedge F$, which serves to cancel the abelian anomaly, 
does arise in this case from the reduction of the eleven dimensional $C \wedge G \wedge G$ term. 
We also  consider the issue of the cancellation of the  
four dimensional anomalies in $M$-theory with a general embedding, and 
discuss its relation to the weakly coupled case.  

Finally, we have discussed  several phenomenological aspects of the behaviour
of the gauge couplings  in  non-standard embedding scenarios. 
The hierarchy of couplings in the visible and hidden sectors is well defined
in the standard embedding.
Changing it,  by which we mean having 
the coupling in the visible sector stronger than in the hidden 
sector (or sectors), can be achieved only by going to 
a non-standard embedding. Such an `inverse' 
scenario has several  advantageous features. One naturally goes around the 
 problem of the existence of a critical radius. Moreover, the condensation
scales become low enough compared
to the hidden sector unification scale  
(since the hidden sector coupling $g_h$ can be made sufficiently small) 
to have a 
suitable hierarchy of masses due to the associated supersymmetry breaking.
Also, 'strong' unification becomes possible.
The hidden sector may have
several components and contains chiral matter, which helps to break 
dynamically supersymmetry in a satisfactory way. 
In addition,  the mixed gauge embeddings with low condensation scale  
revive the 
scenarios where the transmission of the 
supersymmetry breaking is {\em not} gravitational - a situation 
impossible within the restricted framework of the standard embedding models. 

We hope that the basic steps we took in this paper along the route
towards nonstandard embeddings $M$-phenomenology justify further 
investigation of this matter. 

\noindent{\bf Acknowledgements}:

Z.L. and S.P. are supported in part
by the Polish Committee for Scientific Research grant 2 P03B 040 12 1997/98, 
and by the M. Curie-Sklodowska Foundation grant F-115.
The work of S.T. is funded by the Royal Society of Great Britain.
The authors thank John Ellis and Graham G. Ross for interesting 
conversations, and Z.L. would like to thank Stephan 
Stieberger for stimulating discussions.

\end{document}